\documentclass[11pt]{article}
\usepackage{geometry}                
\usepackage{graphicx}
\usepackage{amssymb}
\usepackage{epstopdf}
\title {Integrality theorems in the theory of topological strings}
\author{Albert Schwarz, Vadim Vologodsky}

\begin{document}
\maketitle
\section {Introduction}
The  original goal of this paper was to prove
that the formula for the number of holomorphic
disks given in \cite {AV}, \cite {AKV} always gives an integer (as expected from physical considerations).
Pursuing this goal we simplified and generalized
the proof of integrality of instanton numbers given in \cite {KSV} and \cite {Vol} in such a way that it can be applied also to the situation of holomorphic disks.
We will start with the simplified proof of integrality of instanton numbers and  of some
results of \cite {SV} ; later we will discuss the 
modifications necessary to analyze holomorphic disks.  We have tried to make our paper self-contained; in particular, the paper does not depend on \cite {KSV} and \cite {SV}.

The starting point of the integrality proof
is the consideration of de Rham cohomology
over the ring of integers $\mathbb{Z}$ and over
the ring of $p$-adic integers $\mathbb{Z} _p$.
Defining de Rham cohomology over a ring
we consider differential forms with coefficients
in this ring. The definition can be applied only
to manifold defined over this ring.{\footnote {The
de Rham cohomology of a manifold over a ring $R$ is defined as hypercohomology of a sheaf of differential forms with coefficients in $R$.}} For example,
to define de Rham cohomology over integers
we should work with manifolds over integers.
(In other words, the transition from one coordinate chart to  another one should have
integrality properties, that guarantee that
a form having integer coefficients in one chart has integer coefficients in another chart.)
Correspondingly, the definition of de Rham cohomology over  $\mathbb{Z} _p$ (the ring of $p$-adic integers)  can be
applied only to the manifold defined over this
ring; however, one can prove that for compact
manifolds this cohomology depends only on
corresponding $\mathbb{F}_p$-manifold.
(The canonical homomorphism of  $\mathbb{Z} _p$ onto  $\mathbb{F} _p$, corresponding to the
maximal ideal, generated by $p$, permits us
to consider a manifold over  $\mathbb{Z} _p$
as a manifold over  $\mathbb{F} _p$.) The
map $x\to x^p$ (Frobenius map) is a homomorphism of a field
of characteristic $p$ into itself. It is not a homomorphism of  $\mathbb{Z} _p$, however, it generates a
homomorphism ${\rm Fr}$ (Frobenius transformation) on corresponding cohomology groups.
(The construction of this homomorphism is
complicated, but the origin of it can be traced to  the fact that cohomology 
depends only of  $\mathbb{F} _p$-manifold and the Frobenius map is an endomorphism of
$\mathbb{F} _p$-manifold. See \cite {SS}, Sec. 1.3, for  a short description of the construction of  ${\rm Fr}$.)

Notice that knowing de Rham cohomology over
integers we can calculate de Rham cohomology
over any other ring $R$. This follows immediately from Kuenneth theorem and the remark that the  $R$-module of differential forms over $R$ is a tensor
product of the abelian group of differential forms
over $\mathbb{Z}$ with $R$ . In particular, if there is no torsion
in the cohomology over $\mathbb{Z}$ (or if we neglect torsion) the cohomology over $R$ is
a tensor product of cohomology over $\mathbb{Z}$ with $R$.  This crucial fact permits us to use
Frobenius map on $p$-adic cohomology to obtain information about cohomology over complex numbers. In what follows we assume that the cohomology has no torsion. The inclusion of  the cohomology over $\mathbb{Z}$
into cohomology over any field of characteristic zero specifies an integral structure in the cohomology over a field. In particular, $H_{\mathbb{Q}}(X)$  and $H_{\mathbb{C}}(X)$ have integral structures specified by  $H_{\mathbb{Z}}(X)$. Notice that these integral structures (algebraic integral structures)  are specified only if $X$ is a manifold over integers.
{\footnote { Under certain conditions we can extend our results to the case when the cohomology has torsion. Then we should neglect torsion in cohomology. This means that we consider an
entire ring $R$ without torsion as coefficient ring
and factorize the cohomology with respect
to the torsion subgroup; the quotient will be 
denoted $ H_R(X)$. Notice, that $H_R(X)$ can be
regarded as a subgroup of $H_{F(R)}(X)$ and 
specifies an integral structure in $H_{F(R)}(X)$:
 we say that an element of $H_{F(R)}(X)$ is 
integral if it belongs to this subgroup .  (We use here the notation $F(R)$ for the field of fractions of the ring $R$.)}} 

It is important to emphasize that that there exists
another integral structure  in $H_{\mathbb{C}}(X)$, that can be called topological integral structure. Recall, that for every topological space
$X$ one can define cohomology with coefficients  in any abelian group $G$; we use the notation
$H(X,G)$ for this cohomology. If $X$ is a manifold over integers then $H_{\mathbb{C}}(X)$
is canonically isomorphic to $H(X,\mathbb{C})$;
however, the integral structure in this vector space specified by means of $H_{\mathbb{Z}}(X)$ does not coincide with integral structure
defined by the map of $H(X,\mathbb{Z})$  into
$H(X,\mathbb{C})$ (topological integral structure). Expanding the system of generators in one these integral structures with respect to the generators in another structure we obtain the so called period matrix ; the entries of this matrix are in general transcendental .

 The interplay between these
two integral structures plays an important role in our considerations.

The proof of integrality of mirror map, of instanton numbers, of the number of holomorphic disks consists of two steps. First of all we prove that these  quantities can be expressed in terms of matrix entries of Frobenius map on $p$-adic cohomology. Second, we show that under certain conditions these expressions give $p$-adic integers.  Knowing $p$-integrality for all prime numbers $p$ we can say that the quantities we are interested in are integers; if $p$-integrality can be violated for a finite set of "bad" primes, we can say that in fractions representing the quantities at hand the denominator  contains only "bad" prime factors.

It seems that the first step relating arithmetic geometry to
quantities arising in topological string can be used in both directions: to obtain information about physical quantities from number theory
(as we do in present paper) or to find new ways of dealing with complicated questions of arithmetic geometry starting with ideas of physics. (A small step in the second direction was made in \cite {SS}  where a construction of Frobenius map on $p$-adic cohomology based on the ideas of supergeometry was given.)

The proofs of this paper use many ideas of the 
rigorous proofs in \cite {Vol}. However, we use a different definition of the mirror map; this permits us to weaken conditions of integrality statements. (In our definition we use algebraically integral  sections  instead of topologically integral sections. However, we should require that  algebraically integral sections we are using are related to topologically rational sections.)

Proving $p$-integrality we will always consider
a family of Calabi-Yau manifolds in a neighborhood of a maximally unipotent boundary  assuming that this family is defined over integers and remains smooth after reduction mod $p$; we assume  that the cohomology over $Z_p$ are generated by the 
Calabi-Yau form and its covariant derivatives with respect to an integral coordinate system. 
We give an efficient way to check these assumptions. Under the above assumptions we
prove the $p$-integrality of mirror map. The proof of the $p$-integrality of instanton numbers requires an 
additional condition $p>3$. 

Our proofs are not written in rigorous mathematical way,  but it is not difficult to make rigorous almost all of them (with important
exception of cases when a rigorous proof uses the theory of motives). A more mathematical treatment of main constructions is given in Appendix; a mathematically inclined reader can start with reading it.

\section{Frobenius map on cohomology of Calabi-Yau threefolds}
Let us consider a family of Calabi-Yau smooth threefolds
$X_z$ over a base $\mathcal{M}$. We assume that the
family is 
miniversal.  ( This means that the family contains all complex deformations and the base has minimal possible dimension.) We will say that  the base is the moduli space of complex structures. {\footnote {This terminology is not quite precise; to talk about  moduli space we should consider polarized Calabi-Yau manifolds}}.
We assume that the family is defined over integers;
this means, in particular, that holomorphic
3-forms  $\Omega $ entering the definition of Calabi-Yau manifold have integer coefficients. Moreover, we assume that $\Omega$ is minimal among the forms having these property (i.e. dividing it by non-invertible integer  we obtain a form with fractional coefficients). 

We suppose that the manifolds of our family remain smooth after reduction mod $p$
where $p$ is an arbitrary prime number. 
A rational number can be considered as a $p$-adic number for every $p$ ; to prove that the rational number is an integer it is sufficient to check that the corresponding $p$-adic numbers are integral for all  primes $p$. The assumption of smoothness after reduction mod $p$ can be violated for finite set of prime numbers; then  our methods give $p$-integrality only for the complement of this set. This means that the rational number we consider can have a denominator, but the prime factors of the denominator can belong only to a finite set of "bad" prime numbers.

At first we will consider the family over complex
numbers. Then the cohomology $H(X_z,\mathbb{C})$ constitute a vector bundle over $\mathcal{M}$. {\footnote {More precisely, we should 
regard the family as a map of the union $X$ of $X_z$ onto $\mathcal{M}$ and consider corresponding relative de Rham cohomology.}}This vector bundle is equipped with flat connection $\nabla$ (Gauss-Manin connection).
We will consider the cohomology bundle in the neighborhood of a boundary point $z=0$ of $\mathcal{M}.$  More precisely, we assume that our smooth family over $\mathcal{M}$ can be extended to a non-smooth family over $\bar {\mathcal{M}}$ and $\bar {\mathcal{M}}\setminus {\mathcal{M}}$ is a normal crossings divisor $z^1\cdots z^r=0$.  We assume that the cohomology bundle can be extended to $\bar {\mathcal{M}}$ and the Gauss-Manin connection  has first order poles with nilpotent residues on the extended bundle. {\footnote {These assumptions are satisfied if the family at hand is semistable.}}

We will restrict ourselves to the bundle of
middle-dimensional cohomology  $H^3(X_z,\mathbb{C})$; this is a bundle of symplectic
vector spaces of dimension $2r+2$ where
$r=\dim \mathcal{M}$. (Recall that
the Hodge numbers of Calabi-Yau threefold
are $h^{3,0}=h^{0,3}=1, h^{2,1}=h^{1,2}=r$.)

Let us fix a symplectic basis $g^0, g^a, g_a, g_0$ of flat sections of our bundle  (these sections
are in general multivalued, hence it would be
more precise to consider sections of a vector 
bundle over the universal cover of $\mathcal{M}$). The holomorphic 3-form $\Omega$ can be
written  as linear combination
\begin{equation}
\label{o}
\Omega=X^0g_0+X^ag_a+X_ag^a+X_0g^0
\end{equation}
 Recall that for a given complex structure
 a holomorphic 3-form is defined up to a constant factor. We assume that $\Omega$
 depends holomorphically on complex structure;
 it is defined up to a factor that is holomorphic on $\mathcal{M}.$ The quotients 
 $$t^a=\frac{X^a}{X^0}$$
 can be considered as coordinates on the moduli space $\mathcal{M}$; they are called special
 coordinates. (These coordinates are multivalued  in general, but for  the case we are interested in the coordinates $q^a=\exp{t^a}$ are one-valued and can be extended to the neighborhood of the point $z=0$ in $\bar {\mathcal{M}}$. This point corresponds to $q=0.$ )
 
 One can normalize the form
 $\Omega$ to obtain
 $X^0=1$; then the normalized form $e_0=
 \frac {\Omega}{<g^0, \Omega>}$ is equal to
 $$e_0=g_0+t^ag_a+\frac{\partial f_0}{\partial t^a}g^a-
(2f_0-t^a\frac{\partial f_0}{\partial t^a})g^0,$$\\
where $f_0$ stands for genus zero free energy.
(See, for example, \cite {BCOV} or \cite {ST}).{\footnote {
  To obtain this formula from mathematical viewpoint one should work
on the extended moduli space (moduli space of
pairs (complex structure, Calabi-Yau form)).
The Calabi-Yau forms $\Omega$ and their multiples specify  a Lagrangian cone in the extended moduli space; the free energy $F_0(X^0,X^a)$ is defined as the generating function of this cone. It is a homogeneous function of degree 2; we represent it as $(X^ 0)^2 f_0 (\frac{X^a}{X^0}).$}}

One can include $e_0$ into symplectic basis
$e^0,e^a,e_a,e_0$ where
$$e^0=g^0,\\
e^a=g^a-t^ag^0,\\
e_a=\nabla_ae_0,$$\\
Using that $g^0, g^a, g_a, g_0$ are flat (covariantly constant with
respect to the Gauss-Manin connection $\nabla_a$), we see that
\begin{equation}\label{eq:gauss-manin-conn}
\begin{array}{lll}
&\nabla_b e_0=e_b,\ \ \ \ \ \ & \nabla_b e_a=C_{abc} e^c\\
&\nabla_b e^a=\delta_{ab}e^0,\ \ \ \ \ \ & \nabla_b e^0=0,
\end{array}
\end{equation}
where $C_{abc}=\partial_a\partial_b\partial_c f_0$ are called Yukawa couplings. The Yukawa couplings are holomorphic with respect to $q^a=\exp{t^a}$; the holomorphic part of free energy
can be expressed in terms of them.
Notice, that one can add a polynomial of degree $\leq 2$  with respect to $t^a=\log q^a$ to the free energy without changing any physical quantities; this addition corresponds to the change of the basis $g^0, g^a, g_a, g_0.$ We will assume that $f_0$ contains only terms of degree 3 with respect to $t.$

Let us consider the family at hand in the neighborhood of maximally unipotent boundary point of the moduli space $\mathcal{M}$ of complex structures. Recall, that the coordinates in the neighborhood of the boundary point are denoted by $z^1,...,z^r$; the points on the boundary divisor  obey $z^1\cdots z^r=0$. 

One classifies flat sections according their
behavior at the boundary point $z=0$. Our
assumption that the point is maximally unipotent
means that there exists one (up to a constant
factor) holomorphic flat section, $r$ linearly independent flat sections with logarithmic 
behavior at  $z=0$ (or, more precisely, $r+1$ 
flat sections with at most logarithmic singularities),$r$ flat sections behaving as
$\log ^2$ and one flat section behaving as $\log ^3$. Instead of flat sections one can talk about
solutions of Picard-Fuchs equations. (Recall, that  the Picard-Fuchs equation is the equation for periods, i.e. for expresions  $<g,\Omega>$ where $g$ is a flat section.) Notice, that  under our conditions the holomorphic solution  $<g^0,\Omega>$ of the
Picard-Fuchs equation can be written as a series
with integer coefficients and the constant term
equal to 1. This can be checked directly in all
interesting cases (for example, for hypersurfaces in toric varieties this follows from expressions given in \cite {B}, see also \cite {CK}). The general proof is given  in Appendix.

One can expand flat sections into series with
respect to  $z^i, \log z^i$;  we can consider
filtration  on the space of flat sections $W^0\subset W^1\subset W^2\subset W^3$ ( no logarithms, at most linear in $\log z^i$, at most quadratic in logarithms, etc). This filtration can  be extended to the space of all sections using the remark that every section can be represented as a linear combination of flat sections with variable coefficients. It can be defined also in terms of 
monodromy operators $M_i$ and their logarithms $N_i$; it is called  monodromy weight
filtration. (Notice, that the standard definition of
monodromy weight filtration differs slightly of
the above construction and our notations are not standard. See Appendix.)

We will take the symplectic basis  $g^0, g^a, g_a, g_0$ of flat sections
in such a way that $g^0 \in W^0, g^a\in W^1,
g_a\in W^2, g_0\in W^3$.  More precisely, we
assume that
$$g^a=g^0\log  z^a + h^a$$
where $g^0$ and $h^a$ are holomorphic at $z=0$. Then the special 
coordinates
$$t^a=\frac{<g^a,\Omega>}{<g^0,\Omega>}=
\log z^a+\frac{<h^a,\Omega>}{<g^0, \Omega>}$$
 are called canonical coordinates.
 The coordinates $t^a$ are multivalued;
 therefore instead of them it is convenient to
 use coordinates
 $$q^a=\exp {t^a}=z^a\exp {\frac{<h^a,\Omega>}{<g^0, \Omega>}}$$
 that are also called canonical coordinates.
 The expression of canonical coordinates
 $q^a$ in terms of original coordinates $z^a$
 is called mirror map.
 Notice,  that the definition of canonical
 coordinates $q^a$  specifies them only
 up to  constant factors.   (The canonical
 coordinate $t^a$ does not depend on the choice of $g^0$, however, we have a freedom to
 add a multiple of $g^0$ to $g^a$; hence
 $t^a$ is defined up an additive constant.)
 We will assume that $q^a$ is expressed in terms of
 $z^a$ as a series with rational coefficients. (It is sufficient to assume that the terms linear with respect to $z^a$  have rational coefficients; then solving recursively the equations for the flat section we obtain that all other coefficients are also rational.)
 
 The assumption that the family at hand is defined over integers can be used to fix canonical coordinates and mirror
 map uniquely . We require that $q^a$
 behaves like $z^a$ as $z\to 0$; this means that
\begin {equation}
 \label {h}
 \exp {\frac{<h^a,\Omega>}{<g^0, \Omega>}}=1
 \end {equation}
  at $z=0$. (More generally one can assume that $q^a$ behaves like $\pm z^a$ at $z=0$;   in other
 words $\exp {\frac{<h^a,\Omega>}{<g^0, \Omega>}}=\pm 1$ at $z=0$). {\footnote
  {Notice, that there exists another approach to fixing of mirror map based on consideration topologically integral sections $g^0,g^a.$ Our requirement agrees with the definition
  of monomial-divisor mirror map for hypersurfaces in toric varieties. It is important to emphasize that a toric variety can be considered naturally as a variety over $\mathbb{Z}$; the same is true for hypersurfaces therein if the coefficients of defining equation are integral.}} We assume that  the coordinates $z^1,...,z^r$ agree with integral structure on the moduli space $\mathcal{M}$. It is natural to conjecture that canonical coordinates $q^1,...,q^r$ also have this feature; in other words, the coefficients of expansion of these coordinates in terms of  $z^1,...,z^r$ are integers 
(recall that we assumed that the linear terms
of these series are equal to $\pm z^a$). This
conjecture (integrality of mirror map) was proven in many cases; see
\cite {LY},\cite {KR},\cite {KR1}, \cite {Vol} and the discussion in Sec. 4.  

{\it In the present  section  and in Sec. 3 we will freely use the integrality of mirror map proving integrality of other objects; in Sec. 4
 we will check the integrality of mirror map in the conditions we need.}  

It follows from integrality properties of holomorphic period
$<g^0,\Omega>$  that $e_0=
 \frac {\Omega}{<g^0,\Omega>}$  is an integral vector in algebraic sense (or more precisely, an integral section of the cohomology bundle). Using integrality of mirror map we can see that $e_a$ also has this feature. (The covariant derivatives $\nabla _a$ with respect to canonical
 coordinates preserve algebraic integrality.) The
integrality of normalized Yukawa couplings $C_{abc}$ follows from the formula $$C_{abc}=<e_0,\nabla _a \nabla _b \nabla _c e_0>=
 <\nabla _be_a,e_c>.$$ {\it We will assume that $e^c$ can be expressed as a linear combination of $\nabla _be_a$ with integer coefficients; hence it is an integral vector.} This means, in particular, that  for fixed $c$ the numbers $C_{abc}(0)$ (classical limits of Yukawa couplings) are relatively prime.
 If we are interested in $p$-integrality we should work with $\mathbb{Z}_p$-cohomology; the above assumption takes the form : $C_{abc}(0)$ reduced mod $p$ is a matrix of rank $r$
 (we consider the pair $a,b$ as one index). {\footnote {See Appendix for more mathematical  form of this assumption. }}

We see that the basis 
$e^0,e^a, e_a, e_0$ consists of integral vectors (in algebraic
sense); moreover, these vectors generate the
de Rham cohomology over integers (this follows immediately from the fact that the basis is symplectic). However,  in present
 section we need only
much weaker statement that this basis  consists of algebraically rational vectors.  This statement follows from the rationality condition used in the construction of canonical map.

{\it We will impose also an additional condition
that the sections $g^a$ are topologically rational.} (Recall that topologically rational sections are always flat.) One can check that this condition is satisfied for hypersurfaces in toric varieties.
 
 We assumed that the family at hand is defined over integers; this means that we can consider it
over any other ring, in particular, over $p$-adic
integers $\mathbb{Z}_p$ or over the field of
$p$-adic rational numbers $\mathbb{Q}_p$. (If our family were defined over $\mathbb{Q}$ we still would be able to consider it over $\mathbb{Q}_p$.)
Again we obtain a bundle of corresponding
cohomology with flat connection (Gauss-Manin
connection). {\footnote {Due to the fact that we are working
with cohomology factorized with respect to torsion this statement is not trivial; moreover, it is not always true. However, it was proved in \cite {F} that this statement is true in the case if the family at hand has a smooth reduction mod $p$
and $p>3.$ We always assume that the $\mathbb{Z}_p$-cohomology form a bundle; then one can construct the Gauss-Manin connection on this bundle.}}  It  is important
that Gauss-Manin connections for different rings are compatible. (A homomorphism of rings induces a homomorphism of de Rham cohomology, commuting with connection.) 

We neglect torsion in cohomology,
therefore we can obtain de Rham cohomology with coefficients in
any ring $R$ tensoring the cohomology over
integers with $R$; this operation is compatible with Gauss-Manin connection. Analogously,
the natural map of the cohomology with coefficients in $\mathbb{Q}$ into cohomology
with coefficients in $\mathbb{Q}_p$ commutes
with Gauss-Manin connection.
This means, in particular, that we can consider
$g^0,g^a, g_a, g_0$ and $e^0,e^a,e_a,e_0$ as
bases in $p$-adic cohomology; they are defined in 
a neighborhood of the point $z=0$. {\footnote { More precisely one should consider these bases in a formal neighborhood of $z=0$. This means that we consider them as formal power series with respect to $z^a, \log z^a$.}}  (Recall that $\log$
makes sense as a one-valued function defined for any non-zero $p$-adic number.)

In the neighborhood of $z=0$ we consider a map
$z\to z^p$ transforming every coordinate to its
$p$-th power: $(z^1,...,z^r)\to ((z^1)^p,...,(z^r)^p$.
In the case when $R$ is the ring of integer $p$-adic numbers $\mathbb{Z}_p$ we can  lift the
map $z\to z^p$ to the map ${\rm Fr}$
on  cohomology $H_{\mathbb{Z}_p}(X_z)$. This map (relative Frobenius map) is compatible with Gauss-Manin connection and with the scalar product on cohomology:
\begin{equation}
\label{gmf}
\nabla _a{\rm Fr}=p{\rm Fr}\nabla _a ,
\end{equation}
\begin{equation}
\label{sc}
<{\rm Fr}x,{\rm Fr}y>=p^3{\rm Fr}<x,y>.
\end{equation}
Where we  denote by $\nabla _a$ the covariant derivative on the bundle,
corresponding to the logarithmic derivative
$\delta _a=z^a\frac{d}{dz^a}$ in the neighborhood of $z=0$. 
We consider $x$ and $y$  as sections of the bundle of cohomology; hence $<x,y>$ is a function on the neighborhood of $z=0$.   The action of ${\rm Fr}$ on a function $f(z)$ transforms it into $f(z^p)$.

Notice, that Frobenius map is not linear in standard sense: if $x$ is a section of cohomology bundle and $f$ is function on the base then
\begin{equation}
\label{fr}
{\rm Fr}(fx)= {\rm Fr}(f){\rm Fr}(x).
\end{equation} 

Frobenius map is compatible with monodromy
weight filtration:
$${\rm Fr}W^a\subset W^a.$$
(see Appendix).
The map ${\rm Fr}$ can be extended to the
vector spaces $H_{\mathbb{Q}_p}(X_z)$. Fixing
 bases $e_a(z)$ in these vector spaces we can talk about corresponding matrices $m_a^b(z)$.
 If for  $z\in \mathbb Z _p$ the basis $e_a(z)$ can be considered as a set of generators of   $H_{\mathbb{Z}_p}(X_z)$ then the  matrix has  integral entries (entries
 belonging to $\mathbb Z_p$ ). Moreover, these entries have some divisibility properties; this follows from the fact that applying the Frobenius map to $\Omega$ ( to the cohomology class corresponding to the Calabi-Yau 3-form ) we 
 obtain  a cohomology class divisible by $p^3.$
 If $F^k$ denotes the Hodge filtration one can prove that applying the Frobenius map to a cohomology class belonging to $F^k$ we obtain
 a class divisible by $p^k$. To understand these
 statements one should notice that the Hodge filtration is related to the number of differentials
 $dz$ and $d(z^p)=pz^{p-1}dz$ is divisible by $p.$ However, this remark is a far cry from a proof;
 this is clear from the fact that 
 in general the divisibility properties are valid only
 for $p> 3$. (See \cite {SS}, Sec. 5 , for the proof of divisibility.)  Notice, however, that in the case $k=1$ the condition $p>3$ is not necessary for
 divisibility by $p$. {\footnote {The condition $p>3$ arises because we are working with threefolds. On $n$-dimensional manifolds one should require $p>n.$}} (We will use this fact in Sec.  4.)
 
  If  the elements of the basis 
 are flat sections of the bundle of cohomology
 it follows immediately
 from  (\ref {gmf}) that the entries of the matrix $m_a^b(z)$ are constant (do not
 depend on $z$).
 
  Notice, that the Frobenius map depends on the choice of the coordinate system on the moduli space.  If the coordinate system behaves like $z$
  at the point $z=0$ (more precisely, the Jacobian matrix of the transformation of $z^1,...,z^r$ to the new coordinates is non-degenerate at $z=0$) one
  can say that the Frobenius map is regular at $z=0$ (the matrix entries are holomorphic at this point). In particular, this is true in canonical coordinates.

  Let us  consider  the simplest situation when
 the three-dimensional Betti number is equal
 to four; then the moduli space of complex
 structures on Calabi-Yau manifold is one-dimensional ($r=1$). We will regard $z$ as a coordinate on a punctured disk belonging to the moduli space; the logarithmic derivative with respect to
 $z$ will be denoted by $\delta$ and the corresponding Gauss-Manin covariant derivative by $\nabla$. In the situation at hand cohomology
 classes $\Omega _i=\nabla ^i \Omega$
 for i=0,1,2,3 constitute a basis in three-dimensional cohomology; hence
 \begin{equation}
\label{ }
\nabla  ^4 \Omega=c_3 \nabla  ^3\Omega+c_2 \nabla  ^2 \Omega+c_1 \nabla  \Omega +c_0 \Omega.
\end{equation}
 
 It follows from this formula that for any flat section $g$ the scalar product $<g, \Omega> $ (period) obeys Picard-Fuchs equation 
 \begin{equation}
\label{ }
{\delta} ^4<g, \Omega>=c_3 {\delta} ^3<g,\Omega>+c_2 {\delta} ^2 <g,\Omega>+c_1 {\delta}<g, \Omega> +c_0 <g,\Omega >.
\end{equation}
 
 If all coefficients $c_i(z)$ vanish at the point $z=0$ then this point is a maximally unipotent boundary point.

 Notice that Picard-Fuchs equation has the same form over any field of characteristic zero.

  It is easy to relate the basis $g^0, g^1,g_1,g_0$ to the basis $e^0,e^1,e_1,e_0$ and
 to the 
 basis $\Omega _i$, $i=0,1,2,3.$ The entries of
 the matrix of scalar products of $g^0, g^1,g_1,g_0$ with $\Omega _b$ (period matrix) are obtained from periods  by
 means of differentiation: if $g$ is a flat section then
 $$<g,\Omega _b>=\delta ^b <g, \Omega>$$.
 
 We have seen that
 \begin{equation}
\label{eg}
e^0=g^0,\\\\
e^1=g^1+tg^0,
e_1=g_1+f''_0g^1-(f'_0-tf''_0)g^0,
e_0=g_0+tg_1+f'_0g^1-(2f_0-tf'_0)g^0.
\end{equation}
Where $'$ denotes the derivative with respect to $t$ (=logarithmic derivative with
respect to the canonical coordinate $q=e^t$).

Conversely,
\begin{equation}
\label{ge}
g^0=e^0,g^1=e^1-te^0,g_1=e_1-f''e^1+f'e^0,
g_0=e_0-te_1-(f'-tf'')e^1+(2f-tf')e^0.
\end{equation}

In canonical coordinate the Gauss-Manin connection has the form
\begin{equation}\label{eq:gauss-manin}
\begin{array}{lll}
&\nabla e_0=e_1,\ \ \ \ \ \ & \nabla e_1=Y e^1\\
&\nabla e^1=e^0,\ \ \ \ \ \ & \nabla e^0=0,
\end{array}
\end{equation}
where $Y=\delta ^3 f_0$ (Yukawa coupling) is the third derivative of the free energy $f_0$ with
respect to $t=\log q$ (third logarithmic derivative
with respect to $q$) and $\nabla$ is covariant derivative corresponding to the logarithmic derivative $\delta$.

We will consider more general situation when
the Gauss-Manin connection in symplectic basis
has the form 
\begin{equation}\label{eq:gauss-man}
\begin{array}{lll}
&\nabla e_0=Y_3e_1,\ \ \ \ \ \ & \nabla e_1=Y_2 e^1\\
&\nabla e^1=Y_1 e^0,\ \ \ \ \ \ & \nabla e^0=0.
\end{array}
\end{equation}
It is easy to check that $Y_1=Y_3$. (This follows from the assumption that the basis at hand is symplectic.)

 Let us denote by $m_{a,b}$ the entries
 of the matrix of Frobenius map in the basis
 $e^0,e^1,e_1,e_0$. 
 \begin{equation}
\label{e}
{\rm Fr}e^0=m_{1,1}e^0,
{\rm Fr}e^1=m_{2,2}e^1+m_{1,2}e^0,
{\rm Fr}e_1=m_{3,3}e_1+m_{2,3}e^1+m_{1,3}e^0,
{\rm Fr}e_0=m_{4,4}e_0+m_{3,4}e_1+m_{2,4}e^1+m_{1,4}e^0.
\end{equation}
(The matrix is triangular, because the Frobenius map is compatible with the monodromy weight filtration.)

From the equation (\ref {gmf}) we derive the 
equations for the matrix elements of Frobenius operator:
\begin{equation}
\delta m_{i,i}= 0 
\end{equation}
\begin{equation}
\delta m_{i,i+1}=pm_{i ,i}{\rm Fr}Y_i- m_{i+1,i+1}Y_i
\end{equation}
\begin{equation}
\delta m_{i,i+2}=pm_{i,i+1}{\rm Fr} Y_{i+1}-m_{i+1,i+2}Y_i
\end{equation}
\begin{equation}
\delta m_{1,4}=pm_{1,3}{\rm Fr}Y_3-m_{2,4}Y_1.
\end{equation}
It follows from the first equation that the diagonal entries are constant.
 The matrix entries are holomorphic at $z=0$, hence the LHS of the above equations vanishes.
 We obtain that 
 $$pm_{i,i}=m_{i+1,i+1},$$
 hence
 $$m_{i,i}= p^{i-1}m_{1,1}.$$
 Using (\ref {sc}) we conclude that
 $$m_{1,1}=\pm 1.$$
 In what follows we will assume that $m_{1,1}=1$; the modifications necessary in the case when the entry $=-1$ are 
obvious.
Using vanishing of LHS and (\ref {sc}) we can express  all matrix entries of Frobenius map at
$z=0$ in terms of $\alpha = p^{-1}m _{1,2}(0)$ and $\beta =p^{-3}m_{1,4}(0).$ In particular, $m_{2,3}(0)=
p^2c_0\alpha$ where $c_0=Y_2(0)$,$m_{3,4}=p^3\alpha$, $m_{1,3}(0)=\frac{1}{2}p^2 c_0 \alpha ^2$,
$m_{2,4}(0)= \frac{1}{2}p^3 c_0 \alpha ^2$.  Solving differential equations
we can find the dependence   of matrix entries of $z.$ 

In canonical coordinates (i.e. when $Y_1=Y_3=1$) the solutions
have the following form:
 $$m_{1,2}=\alpha,$$
$$m_{2,3}=({\rm Fr} f''_0p-f''_0p^2)+p^2 c_0\alpha$$
$$m_{1,3}=-({\rm Fr} f'_0-p^2 f'_0)+\frac{1}{2}p^2 c_0 \alpha ^2+{\rm Fr}{\tilde f}''_0p\alpha$$
$$m_{3,4}=p^3\alpha$$
$$m_{2,4}={(\rm Fr}f'_0p-p^3f'_0)+ \frac{1}{2}p^3 c_0 \alpha ^2-p^3\alpha {\tilde f}''_0$$
$$m_{1,4}=-2({\rm Fr}f_0-p^3f_0)+
p^3\beta+
p \alpha {\rm Fr}{\tilde f}'_0 -p^3\alpha {\tilde f}'_0.$$
Here ${\tilde f}_0$ stands for the holomorphic  (with respect to $q$) part of free energy. Recall, that  the third derivative of $f_0$ is
  Yukawa coupling $Y=\sum _{k\geq 0} c_kq^k$, hence we can take
  $$f_0=c_0\frac{t^3}{3!}+{\tilde f}_0$$
   where $$ {\tilde f}_0=\sum _{k>0}\frac{c_k}{k^3}q^k.$$
       
 There exists a more direct way to obtain these formulas using the basis consisting of flat connections.     
      
We denote by $\mu _{a,b}$ the entries of the matrix of Frobenius map in the basis
 $g^0, g^1,g_1,g_0$.
 \begin{equation}
\label{g}
{\rm Fr}g^0=\mu _{1,1}g^0,
{\rm Fr}g^1=\mu_{2,2}g^1+\mu_{1,2}g^0,
{\rm Fr}g_1=\mu_{3,3}g_1+\mu_{2,3}g^1+\mu_{1,3}g^0,
{\rm Fr}g_0=\mu_{4,4}g_0+\mu_{3,4}g_1+\mu_{2,4}g^1+\mu_{1,4}g^0.
\end{equation}

   This matrix is
 triangular because ${\rm Fr}$ is compatible with
 monodromy weight filtration; it   is a constant matrix because the basis
 consists of flat sections.

Using the relations (\ref {eg})  and (\ref {ge})
we can express the matrix $m_{a,b}$ in canonical coordinates in terms of 
the constant matrix $\mu _{a,b}$.

We obtain
$$m_{i,i}=\mu _{i,i}$$
$$m_{1,2}=\mu _{1,2},$$
$$m_{2,3}=({\rm Fr} f''_0p-f''_0p^2)+\mu _{2,3}$$
$$m_{1,3}=-({\rm Fr} f'_0-p^2 f'_0)+p^2\mu _{2,3}t+\mu _{1,3}+{\rm Fr}f''_0p\mu _{12}$$
$$m_{3,4}=\mu _{34}$$
$$m_{2,4}={(\rm Fr}f'_0p-p^3f'_0)+\mu _{2,4}+t\mu _{2,3}-\mu _{3,4}f''_0$$
$$m_{1,4}=-2({\rm Fr}f_0-p^3f_0)+
\mu _{1,4}+t\mu_{2,4}+t\mu _{1,3}p+
t^2\mu _{2,3}+{\rm Fr}f'_0\mu _{1,2}-f'_0\mu _{3,4} $$
 It follows from these formulas that 
 $$\mu _{i,j}=m_{i,j}(0).$$
 
 Notice, that in the derivation we used the assumption that $f_0$ contains only cubic term with respect to $t.$
 
  As we mentioned the terms in $m_{a,b}$ that contain $t=\log q$ should vanish. This means that we can  replace everywhere $f_0$  with its holomorphic (with respect to $q$) part ${\tilde f}_0$ and omit all terms with explicit dependence of $t$ . This leads to the formulas we obtained solving differential equations. ( In some expressions containing
  $f_0$ nonholomorphic terms cancel , hence
  there is no necessity to replace $f_0$ with its holomorphic part.)
  
 The calculation of  $m_{i,j}(0)$ (of the behavior of  entries of Frobenius matrix at the maximally unipotent boundary point) is much more difficult.
 It is based on relation between flat sections  $g^0$,
 $g^1$  that {\it we assume to be topologically rational} with algebraically rational sections $e^0, e^1$.  If we are working in a
 coordinate system $z$ compatible with integral structure of family of varieties at hand one can
 prove that $m_{1,2}(0)=0,$ hence  $m_{2,3}(0)=0$, $m_{3,4}(0)=0$ . (For the case of quintic this can be proved by means of direct calculation \cite {q}.)  If coordinate $u$ is expressed in terms of  $z$ by a series  with rational coefficients and the series starts with $cz$ (i.e. $u$ is compatible with the structure of the family over $\mathbb{Q}$ and in the neighborhood of the boundary point $u$ behaves like $cz$) then the matrix element $m_{1,2}(0)$ of the Frobenius operator calculated in the coordinate $u$ is equal to $Y_3(0)\log (c^{p-1})$ where the logarithm is understood in $p$-adic sense.
 
 The proof is based on the
  theory of motives. (The theory of motives can be regarded as a kind of "universal cohomology theory".)  We will skip the proof referring to  \cite {Vol}, but we will mention some salient steps. The main point is a construction of a motive over $\mathbb{Q}$ that has  realizations as 
complex Hodge structure  and "$p$-adic Hodge structure" (Fontaine-Laffaille structure).  (In both cases we have in mind the limiting Hodge structures, i.e. Hodge structures over the boundary point.) The $p$-adic Hodge structure 
has Frobenius map as one of ingredients, hence the motive can be used to relate the matrix of Frobenius to the period matrix over complex numbers. 

 The general statement  looks as follows.

Let  us consider a symplectic rational  basis  $e^0,e^1,e_1,e_0$ where $e_0$ is a Calabi-Yau form and
the Gauss- Manin connection is given by the
formula (\ref {eq:gauss-man}) where covariant derivatives correspond to logarithmic derivatives 
 with respect to the coordinate $u.$ Let us assume
that $Y_3(0)=1.$ Then at
$u=0$  the superdiagonal entry of the matrix of periods has the form
 $\frac{1}{2\pi i}\log c$ 
where $c$ is a product of a rational number and a root of unity. {\footnote {This means that for topologically rational logarithmic section $g^1$  the period $<g^1, e_0>$ behaves like $ \frac{1}{2\pi i}(\log u+\log c)$ as $u\to 0.$ (We normalize $g^1$ imposing the condition $<g^1, e_1>=1.$) Notice,
that $g^1$ is defined up to a summand $\alpha g^0$ where $g^0=e^0$ is a topologically (and algebraically) integral holomorphic section and $
\alpha$ is a rational number. Hence $c$ is defined only up to a factor $\exp (2\pi i \alpha)$.
This factor is a root of unity, $p$-adic logarithm of it is equal to zero, hence it does not contribute to  the expression for $m_{1,2}(0).$ }}The  entry  $m_{1,2}(0)$ of Frobenius matrix in the coordinate $u$ can be expressed in terms of $c$ as $\pm \log c ^{1-p}$ where $c$ should be understood as  a $p$-adic number.

In canonical coordinates $c=1$, hence $m_{1,2}(0)=0.$ We obtain the same answer for this entry 
in any integral coordinate system, because the answer depends only on behavior of coordinates at the boundary point.


Notice, that as long we are interested  in  the behavior of  Frobenius operator only at boundary point it is sufficient to calculate this behavior   in one coordinate system, because there exists a general formula relating the matrices of Frobenius operators in  different coordinates
(see \cite {Vol}).

 Applying  the above consideration to the matrix of Frobenius operator in canonical coordinate $q$ we obtain that, in particular,
 \begin{equation}
\label{14}
m_{1,4}=-2({\rm Fr}f_0-p^3f_0)+
p^3\beta.
\end{equation}

The proof of (\ref {14}) was given in the case
when the dimension $r$ of the moduli space of
complex structures is equal to $1$. 
If $r>1$ we can calculate  the matrix of Frobenius map in symplectic basis $e^0, e^a,e_a,e_0$ of the space of sections of the cohomology bundle if
the Gauss-Manin connection in this basis
has the form 
\begin{equation}
\label{gm}
\begin{array}{lll}
&\nabla _a e_0=(Y_3)_a^{b}e_b,\ \ \ \ \ \ & \nabla _ae_b=(Y_2)_{abc} e^c\\
&\nabla _a e^b=(Y_1)_a^b e^0,\ \ \ \ \ \ & \nabla _ae^0=0.
\end{array}
\end{equation}
where $Y_1=Y_3$.

Canonical coordinates are defined by the condition  $(Y_1)_a^b=(Y_3)_a^b= \delta _a^b$.

Basically the same
arguments as for $r=1$ can be used to prove (\ref {14}) in canonical coordinates in
the basis $e^0,e^a,e_a,e_0.$

Notice that  $(Y_1)_a^b=(Y_3)_a^b$ coincides with $y_a^b$ introduced  in Sec. 4 where we prove $p$-integrality of the mirror map under the condition that  at the boundary point the reduction of this matrix mod $p$ is nondegenerate.  It is important to emphasize that this condition follows from our assumptions about normalized Yukawa couplings reduced mod $p$. To check this we consider non-normalized Yukawa couplings defined by the  formula
$<e_0,\nabla _a \nabla _b \nabla _c e_0>$
where the covariant derivatives are taken in the original coordinate system (as in (\ref {gm})). At the boundary point canonical coordinates coincide with the original ones, hence the difference between normalized and non-normalized Yukawa couplings disappears.
From the other side using (\ref {gm}) we can express non-normalized Yukawa couplings in terms of   $Y_i$; it follows immediately from this expression that for degenerate $y_a^b(0)$ mod $p$ our assumptions about Yukawa couplings are violated.

In the next section we will use (\ref {14})       to express the instanton numbers in terms of Frobenius map.

\section {Instanton numbers}

In the case when there is one Kaehler parameter 
(in terms of A-model) or the moduli space of complex structures on Calabi-Yau manifold is one-dimensional  (in mirror B-model) the instanton numbers $n_k$ are related to the genus zero free energy $f_0$ 
  in the following way:
\begin{equation}
\label{i}
{\tilde f}_0(q)=\sum _{k=1}\sum _{d=1}d^{-3}n_kq^{dk}=\sum _{k=1}n_k{\rm Li}_3(q^k)
\end{equation}
where ${\tilde f}_0$ stands for the holomorphic part of free energy and $q$ denotes the canonical coordinate. (We use here the notation
${\rm Li}_k(x)=\sum_ {s=1}  s^{-k}x^s.$)

Equivalently one can relate the instanton numbers to the (normalized) Yukawa coupling:\begin{equation}
\label{y}
Y(q)=const +\sum _{k=1} n_kk^3\frac{q^k}{1-q^k}=const+\sum _{k=1}\sum _{d=1}n_kk^3q^{dk}
\end{equation}

The following lemma  permits us to use $p$-adic methods in the analysis of integrality of instanton numbers.

{\bf Lemma 1.} The numbers $n_k$ defined in terms of $Y(q)$ by the formula (\ref {y}) are $p$-adic integers if and only if   there exists such a series $\psi (q)=\sum s_kq^k$ having $p$-adic integer coefficients that
$${\rm Fr}Y-Y=\delta ^3 \psi$$
(or, in other words, $Y(q^p)-Y(q)=\delta ^3 \psi (q).$  Hence the numbers $n_k$ are integers if
this condition is satisfied for every $p$.

We will give a proof  of this lemma based on the notion of Dirichlet product of arithmetic functions.(Another proof that does not use Dirichlet product is given in \cite {KSV}.) Recall, that an arithmetic function is a function defined on the set $\mathbb{N}$ of natural numbers and taking values in complex numbers (or in any other commutative ring).
The Dirichlet product of two arithmetic functions is defined by the formula
\begin{equation}
\label{dir}
(f\ast g)(s)=\sum _{dk=s}f(d)g(k)=\sum _{d|s}f(d)g(\frac{s}{d})
\end{equation}
Where $d,k,s$ are natural numbers, $d|s$ means that $d$ divides $s$.

Arithmetic  functions form a commutative, associative unital ring with respect to Dirichlet product.

The relations (\ref {i}) and (\ref {y}) can be written as
$$({\tilde f}_0)=\{\frac{1}{k^3}\}\ast\{n_k\},$$
$$(Y)=\{1\}\ast\{n_kk^3\}.$$
Here we use the notation $\{c_k\}$ for the arithmetic function, corresponding to the sequence $c_1,...,c_k,...$ and the notation $(f)$
 for the arithmetic function, corresponding to the sequence of coefficients of power expansion of $f$. (The equivalence of (\ref {i}) and (\ref {y})
 follows from the remark that $\delta ((f)\ast(g))=
 \delta (f)\ast\delta (g)$.)

Using the new form of (\ref {y}) we can express
the instanton numbers in terms of Yukawa coupling:
\begin{equation}
\label{n}
{n_kk^3}=\mu \ast (Y).
\end{equation}
 Where $\mu (k)$ stands for the Moebius function
that is equal to $(-1)^s$ if  $k$ is represented as a
product of $s$ distinct primes and vanishes if
$k$ is not square-free. (This expression follows from the remark that $\mu\ast{1}=(q)$ is the unit  of Dirichlet ring.) 

 The proof of lemma 1 can be based on the following statement 
 
{\bf  Lemma 2.} If $g=\mu \ast (h)$ and $m=({\rm Fr}h-h)$ then
$$g(p^at)=-\sum _{d|t}\mu ({d})m\big({\frac{p^at}{d}}\big).$$
Here $t$ is a natural number that is not divisible by $p$.

Let us denote the set of such numbers by $\mathbb{N}_p$. It easy to check that the definition of Dirichlet product makes sense for functions defined on $\mathbb{N}_p$ and that Lemma 2 can be formulated in terms of this product:
$$g^{(a)}= \mu \ast(h-{\rm Fr}h)^{(a)}$$
where $g^{(a)}$ is a function on  $\mathbb{N}_p$
defined by the formula $g^{(a)}(r)=g(p^ar).$

The proof of Lemma 2 follows immediately from the properties of the Moebius function: we use that
$\mu (pt)=-\mu (t), \mu (p^at)=0 $ for $a>1$.

 Lemma 2 combined with  (\ref {14}) gives an expression of instanton numbers (or, more precisely, of their $p$-adic reduction) in terms of
${\rm Fr}Y-Y$. If ${\rm Fr}Y-Y=\delta ^3 \psi$ where $\psi$ is a series with $p$-adic integral coefficients we obtain $p$-adic integrality of instanton numbers. To prove the remaining statement of Lemma 1 we notice that it follows from
(\ref {y}) that
$${\rm Fr}Y-Y=\sum _{k=1}\sum _{d=1}n_{k}k^3q^{dpk}-\sum _{k=1}\sum _{d=1}n_kk^3q^{dk}.$$
In the RHS the first summand cancels the terms of the second summand that correspond to $d$
divisible by $p$. We obtain that the coefficient in front of $q^t$ is divisible by $ t^3$ in $p$-adic sense. This proves the statement of Lemma 1.

One can reformulate Lemma 2 in the following way

{\bf Lemma 3}.  If $v=\{\frac{1}{k^s}\}\ast g$ (i.e.
$\sum v(k)q^k=\sum _{r=1}g(r){\rm Li}_s(q^r)$)
 and
$m=(p^{-s}{\rm Fr}v-v),$ then
$$ g(p^at)=-\sum _{d|t}\frac{1}{d^s}\mu ({d})m\big(\frac{p^at}{d}\big).$$
Here $v$ and $g$ are arithmetic functions,
 $t$ is a natural number that is not divisible by $p$.

Lemma 3 follows from Lemma 2 applied to
$\delta ^sv=\{1\}\ast \delta ^s g$ and
$\delta ^s m=({\rm Fr}\delta ^s v-\delta ^s v).$

Combining  Lemma 3 with the formula (\ref {14}) of Sec. 2  one can calculate the instanton numbers in terms of the matrix of Frobenius map 
in the basis $(e^0,e^1,e_1,e_0).$  Corresponding expression will be written later in more general situation (see (\ref {in})).

Similar statements can be proved in the case of multidimensional moduli spaces.
In this case instanton numbers specify a
multidimensional arithmetic function $\{n_{\bf k}\}$
where ${\bf k}=(k^1,...,k^r)$ is a multiindex having natural numbers as components; the same is true for the coefficients of power series for ${\tilde f}_0$(
holomorphic part of  free energy).
They are related by the formula
$${\tilde f}_0({\bf q})=\sum _{d|{\bf k}}\frac{{\bf q}^{\bf k}}{d^3}n_{\frac{\bf k}{d}}$$ 
or, equivalently,
 \begin{equation}
\label{ }
({\tilde f}_0)=\{\frac{1}{d^3}\}\ast\{n_{\bf k}\}.
\end{equation} 
We use here the fact that the multidimensional
arithmetic functions constitute a module over
Dirichlet ring: if $g$ is an element of this ring and $h$ is a multidimensional arithmetic function we define 
$$g\ast h= \sum _{d|{\bf k}}g(d)h(\frac{{\bf k}}{d}).$$

Lemma 3 can be generalized immediately to the case of multidimensional arithmetic functions     $v,g,m$. We obtain

{\bf Lemma 3$'$}.  If $v=\{\frac{1}{k^s}\}\ast g$ (i.e.
$\sum v(\mathbf{ k})q^{\mathbf{ k}}=\sum _{r=1}g(\mathbf{ r}){\rm Li}_s(q^{\mathbf{ r}})$)and
$m=(p^{-s}{\rm Fr}v-v),$ then
$$ g(p^a{\bf t})=-\sum _{d|{\bf t}}\frac{1}{d^s}\mu ({d})m\big(\frac{p^a{\bf t}}{d}\big).$$
Here ${\bf t}$ is a multiindex that is not divisible by $p$.

{\bf Corollary.} If $m=(p^{-s}{\rm Fr}v-v)$  has
the form $\sum m({\bf k})q^{\bf k}$ where $m({\bf k})$ are $p$-adic integers, then $v=\sum _{r=1}g(\mathbf{ r}){\rm Li}_s(q^{\mathbf{ r}})$ where
the coefficients $g(\mathbf{ r})$ also are $p$-adic integers. Conversely, the integrality of   $g(\mathbf{ r})$ implies the integrality of  $m({\bf k})$.

   Using Lemma 3$'$ we obtain the expression of
instanton numbers in terms of the coefficients $M_{\bf k}$ of the power series for the entry $m_{1,4}$  of the matrix of Frobenius map in the basis $e^0,e^a,e_a,e_0$. Namely,
\begin{equation}
\label{in}
n_{p^a{\bf k}}=\frac{1}{2}\sum_{d|{\bf k}}\frac{1}{p^3d^3}\mu (d)M_{\frac{p^a{\bf k}}{d}}
\end{equation}
where $\bf k$ is a vector that is not divisible by $p$.

To check the $p$-integrality of instanton numbers it
is sufficient to prove that the coefficients of matrix entry $m_{1,4}$ are integers divisible by $p^3$.
If $p>3$  the integrality of $p^{-3}m_{1,4}$  would follow from the $p$-adic integrality of the basis $e^0,e^a,e_a,e_0$(more precisely, we should prove that the basis  consists of sections that are integral in algebraic sense  and these sections  constitute a
system of generators over $\mathbb{Z_p}).$

We have assumed that $\Omega$ is an integral 
cohomology class; to check that $e_0$ is also 
an integral class we should  use that the holomorphic period (holomorphic solution to the Picard-Fuchs equation) $<g^0, \Omega>$ is a series with respect to $q$ having integer coefficients
and that the zeroth order term in this series is
equal to $\pm 1$. (We have discussed this statement in Sec. 2.) 

 Integrality of vectors $e_a =\nabla _a e_0$
follows from the integrality of the mirror map
(\cite {LY},\cite {KR},\cite {KR1}, \cite {Vol} and the next section), or, equivalently, from the fact that the canonical coordinates $q^a$ are
integral. (It follows  from this fact that covariant
derivatives with respect to these coordinates 
preserve integrality.) The same arguments 
prove that the basis $e^0,e^a,e_a,e_0$ becomes integral
after reduction to $p$-adic numbers if the expression of $e^a$ in terms  of $\nabla _be_c$ is invertible mod  $p$ at $z=0$. {\footnote { One can formulate this condition in more transparent, but less explicit way: the
Calabi-Yau form and its covariant derivatives with respect to coordinate system compatible with integral structure should generate the de Rham cohomology over $\mathbb{Z}_p.$}}  If $r=1$ this means that $p$ does not divide $Y(0)$; then $Y(0)$ is an invertible element of 
of $\mathbb{Z}_p$.  (See Sec. 2 and  Appendix for more detailed discussion of our assumptions about  Yukawa couplings $C_ {abc}(0)=
Y_{abc}(0)$ in the case $r>1.)$  {\footnote {For $p>3$ one can prove the $p$-adic integrality of the basis $e^0,e^a,e_a,e_0$ without
this additional assumption; the proof is based on  theorem by Faltings \cite {F} stating that inner product is a perfect pairing on the
cohomology group  $H_{\mathbb{Z}_p}$.  However, one can conjecture that our assumptions follow from smoothness 
of reduction mod $p$ required in \cite {F}. }}
Notice, that the Yukawa couplings at the boundary point correspond to the structure constants of classical cohomology ring  on
A-model side and can be easily calculated; for quintic $Y(0)=5$.  In Appendix they are interpreted as residues of Gauss-Manin connection on the special fiber.

\section {Integrality of mirror map}

Let us consider the Hodge filtration $F^3\subset F^2\subset F^1\subset F^0$ on the cohomology
with complex coefficients. Recall that
$F^p$ is a direct sum of all groups $H^{k,l}$
where $k\geq p$. We restrict ourselves to the  middle-dimensional cohomology $H^3$, hence
$\dim F^3=h^{3,0}=1, \dim F^2/F^3=h^{2,1}=r,
\dim F^1/F^2=h^{1,2}=r, \dim F^0/F^1=h^{0,3}=1.$ Recall that on the sections of the vector bundle of middle-dimensional cohomology in a neighborhood of the maximally unipotent boundary point we have also the monodromy weight filtration $W^0\subset W^1\subset W^2\subset W^3$ with $\dim W^0=1, \dim W^1/\dim W^0= \dim W^2/W^1=r, \dim W^3/W^2=1.$ One  can represent the space of sections  as a direct sum of  $F^k$ and
$W^{k-1}$; hence this space is a direct sum of
$W^k\bigcap F^k$ where $k=0,1,2,3$;   in particular, $W^1$ is a direct sum of $W^0$ and
$W^1\bigcap F^1$. The quotient $Gr^k_W=W^k/W^{k-1}$
is isomorphic to $W^k\bigcap F^k$.The Gauss-Manin connection descends to this quotient and has trivial monodromy there. We can take a symplectic basis $e^0, e^a,e_a,e_0$ of the space of sections of the cohomology bundle in
such a way that
the Gauss-Manin connection in this symplectic basis
has the form (\ref {gm}).
(To construct this basis one can take flat sections
of $Gr^k_W=W^k/W^{k-1}$ and lift them to
$W^k\bigcap F^k$. The formula for Gauss-Manin connection follows from Griffiths transversality.)
All these statements are proven, for example, in 
\cite {M},[ \cite {D} and \cite {CK}.

 It is a  non-trivial fact, that under certain conditions these statements remain correct also for de Rham cohomology with 
integer coefficients. (This means, in particular, that the entries of the matrix  (\ref {gm}) are series with integral coefficients.) This fact follows from the results of \cite {Vol}. The proof uses reduction to $p$-adic numbers; it gives $p$-integrality for $p>3$ (as always we assume that the family remains smooth after reduction mod $p$). We sketch the proof in Appendix.

To prove the integrality of mirror map we need only a part  of these statements. We will formulate the necessary facts and sketch the proof. The most important step is the expression
of mirror map in terms of the matrix of Frobenius transformation.

To analyze the mirror map we take a basis of $W_1$  consisting of  algebraically integral sections $e^0,e^a$ that can be extended to the boundary point $z=0$ and obey $e^0\in W_0, e^a\in F^1\bigcap W_1$,
\begin{equation}
\label{w}
\nabla _a e^0=0,
\nabla _a e^b=y_a^b(z)e^0.
\end{equation} 
Here the functions $y_a^b(z)$ are holomorphic at $z=0$.
To construct such a  basis  we notice that both 
 $W_0$
and $W_1/W_0$ are trivial bundles; more precisely, they can be characterized
as trivial Hodge structures $\mathbb{Z}$ and $\mathbb{Z}(1)^r$. We define $e^0$ as an algebraically (and topologically) integral generator 
of $W_0$ and obtain $e^a$ by lifting an algebraically integral basis of flat sections
of $W_1/W_0$ to $F^1\bigcap W_1$. 

We  can consider $e^0,e^a$ also as sections of $\mathbb{Q}_p$ cohomology bundle; they constitute a basis in the corresponding version of $W_1$. (These
vectors can be considered also as sections of $\mathbb{Z}_p$- cohomology, but it is not clear whether they generate $W_1$ in this cohomology.)

The matrix of Frobenius operator in the basis
 $e^0,e^a$  has the form
  \begin{equation}
\label{ee}
{\rm Fr}e^0=m_{1,1}e^0,
{\rm Fr}e^a=(m_{2,2})^a_be^b+m_{1,2}^ae^0.
\end{equation}
The entries of this matrix obey equations coming from (\ref {gm})
\begin{equation}
\delta _a m_{1,1}= 0 
\end{equation}
\begin{equation}
\delta _a (m_{2,2})^b_c= 0 
\end{equation}
\begin{equation}
\delta _am_{1,2}^b=pm_{1 ,1}{\rm Fr}y^b_a- (m_{2,2})^c_ay^b_c
\end{equation}
It follows from these equations that the diagonal entries are constant. 
One can prove that $m_{1,1}=\pm 1$, $(m_{2,2})^a_b=\pm p\delta ^a_b.$
This follows immediately from the theory of motives (from the fact that there are
motives with realizations $W_0$ and $W_1/W_0$).  However, it is possible to give a more elementary proof of this fact; it is given in Appendix.
We see that the last equation takes the form
\begin{equation}
\label {12}
\delta _am_{1,2}^b=p{\rm Fr}y^b_a- py^b_a
\end{equation}
(We have assumed that $m_{1,1}= 1$.)

Applying (\ref {12}) we can express the mirror map in terms of the entry $m_{1,2}^b$ of the matrix of Frobenius transformation. We are using in this expression  the Artin-Hasse exponential
function
$$E_p(x)=\exp (x+\frac{x^p}{p}+\frac{x^{p^2}}{p^2}+...).$$
The expansion of this function with respect to $x$ has integer
$p$-adic coefficients: $E_p(x)\in \mathbb{Z} _p [[x]].$
It is easy to check that
$$\delta \log E_p(x^{pk}) -\delta \log E_p(x^k) =-kx^k=-\delta x^k.$$
Using this identity we obtain the following 

{\bf Lemma 4}

Let us suppose that  $r^i(z)\in \mathbb{Q}_p[[z_1,...,z_r]].$ Then for the infinite product
$$Q^i(z)=z_i\Pi_{\bf k}E_p(z^{\bf k})^{-r_{\bf k}^i},$$
where $r_{\bf k}^i$ are coefficients of the power expansion of $r^i(z)$ we have
$${\rm Fr}\nu_j^i -\nu _j^i=\delta _j r^i$$
where
$\nu _j^i =\delta _j \log Q^i$ . If $r^i(z)\in \mathbb{Z}_p[[z_1,...,z_r]]$ then $Q^i(z)\in \mathbb{Z} _p [[z_1,...,z_r]]$ .

Here $\delta_j$ stands for the logarithmic derivative with respect to $z_j$ and the Frobenius map ${\rm Fr}$ transforms $f(z_1,...,z_r)$ into $f(z_1^p,...,z_r^p).$ 
We consider $\bf   k$ as multiindex.

One can give also a different expression for
$Q^i(z)$, namely
$$Q^i(z)=z_i\prod _{\bf k} ( 1-z^{\bf k})^{{s_{\bf k}}^i}$$
where 
$$s_{\bf k}^i=\sum \frac{\mu (d)}{d}r_\frac{\bf k}{d}^i,$$ 
$d$ runs over divisors of $\bf k$, that are not divisible by $p$.

This expression can be obtained from the product formula for Artin-Hasse exponent or directly from Lemma 2 of Sec. 3.

Let us notice now that the mirror map $q(z)$ obeys:
$$\delta _a \log q^b(z)=y_a^b(z).$$
(This follows immediately from the fact that in canonical coordinates $q$ we can write an analog of the formula (\ref {w}) with $y^a_b$ replaced by $\delta ^a_b$.) Taking into account that at $z=0$ the canonical coordinate $q^a$ should behave like $z_a$ we obtain from the Lemma applied to $r^a(z)=\frac{1}{p}m_{1,2}^a$  an expression for the mirror map. To prove
$p$-adic integrality of mirror map it is sufficient to check that $e^0,e^a$ generate $W_1$ in the cohomology with coefficients in $\mathbb{Z}_p.$
(It is well known that  the Frobenius map on $F^1$ is divisible by $p$ .) This can be easily checked if the matrix $y^b_a(0)$ is nondegenerate after reduction mod $p$.  

Notice  that at least for $p>3$ the nondegeneracy of the matrix $y^b_a(0)$ reduced mod $p$ can be derived from the conditions imposed on the classical limit of Yukawa 
couplings in Sec. 2. (See the discussion at the end of Sec. 2. A different form of these conditions is given in Appendix)

The condition of integrality of mirror map can be
expressed also in terms of topologically integral sections.

The space $W^1$ is generated by
topologically integral sections $g^0\in W^0,
g^a\in W^1, a=1,...,r$. Corresponding solutions to the Picard-Fuchs equations have the form $f^0(z)$ and
$f^a(z) = \frac{1}{2\pi i}m^a_b f^0(z) \log z^b+r^a(z)$ where $f^0$ and $r^a$ are holomorphic in the
neighborhood of $z=0$ and $m^a_b$ are integers. (This form of $f^a$ follows from the remark that the monodromy preserves topological integrality.) We will assume that $\det m^a_b=\pm 1$; this equation was called small monodromy condition in \cite {Vol} and Morrison integrality conjecture in \cite {CK}.  If it is satisfied then without loss of generality one can assume that $m^a_b=\delta ^a_b$: one should 
work in another integral coordinate system.
Assuming that $m^a_b=\delta ^a_b$ we define
the mirror map  by the formula  
\begin{equation}
\label{m}
\log q ^a= {2\pi i}\frac{<g^a, \Omega>}{<g^0, \Omega>}
\end{equation}
If the constant term of the series for $q^a$ is integral one can prove $p$-adic integrality of all
coefficients of the series
for such $p$ that our family remains smooth after reduction mod $p$ \cite {Vol}. (It is sufficient to assume that the constant term is rational; than
one can check that  it is an invertible integer.)

 \section{ Counting holomorphic disks.}
 If we consider an A-model on a Calabi-Yau threefold we can try to calculate the number of holomorphic disks with boundary on a Lagrangian submanifold. This problem is ambiguous from mathematical viewpoint (one needs additional information called framing) , but the
 corresponding mirror problem  is well defined
 \cite {AV}, \cite {AKV}.  In the mirror problem we
 should work with $B$-branes. ($B$-branes are represented as vector bundles or, more generally, as complexes of vector bundles or coherent sheaves; in more abstract language they can be considered as
  objects of the derived category on the mirror threefold $X$.) The number of holomorphic disks can be expressed
 in terms of holomorphic Chern-Simons functional. However, if the manifold $X$ is compact it is possible to express this number also in terms of algebraic Chern class $c_2$ (see \cite {MW}). The expressions we will use are based on the ideas of \cite {MW} and of
 \cite {L}, \cite {L1}, \cite {L2}), \cite {L3} .
 
To specify a theory of open strings in the framework of B-model we should fix a $B$-brane
or, in mathematical terms, a complex of vector bundles. (Instead of complexes of vector bundles one can consider coherent sheaves or
algebraic subvarieties.)
 The algebraic Chern class $c_2$  of a complex of vector bundles is an element of the Chow ring;
 it can be represented as a linear combination
 of subvarieties of complex dimension 1. {\footnote {We can define Chern classes considering zero sets of sections of appropriate vector bundles. One can consider these sets as
 topological cycles; then we obtain topological
 Chern classes. Considering these sets  as linear combinations of  algebraic varieties we obtain a definition of algebraic Chern classes.)}} We
 assume that the topological Chern class   vanishes (i.e. the algebraic Chern class is homologous to zero). The number of holomorphic disks can be expressed in terms of the integral of Calabi-Yau 3-form $\Omega$ over a chain having the fundamental class of algebraic Chern 
 class as a boundary. (This integral can be interpreted in terms of complex Chern-Simons
 action; see \cite {MW}.)

 In other words
  the number of holomorphic disks is expressed in terms of potential $\mathcal   
  W$ defined by means of pairing  of the cohomology class of $\Omega$ with an element of the  relative homology group $H_3(X,Y)$, where $Y$ is a (in general disconnected) one-dimensional variety. At first we will consider the case when there are no open string moduli (the  set of $B$-branes is discrete in some sense). The typical situation of this kind was considered in \cite {MW}. We  will work  in a
  neighborhood of the boundary point of $\mathcal{M}$ (of the moduli space of complex structures on $X$) with maximally unipotent  monodromy ) . We normalize  $\mathcal   
  W$ replacing the form $\Omega$ in its definition by the normalized form $e_0$ and express the  normalized potential $\tilde {\mathcal   
  W}$ in terms of 
  canonical coordinates $q=(q^1, ..., q^r)$:
  $$\tilde {\mathcal   
  W}=\sum w_{\bf k}q^{\bf k}+...$$
  Here ${\bf k}$  is a multiindex with non-negative components,  ${\bf k}\neq 0$. We omit logarithmic  terms  in the neighborhood of $q=0$ (notice that the constant term also is omitted).   Of course, the normalized potential depends on
  the choice of relative cohomology class paired with $e_0$; this means that it is defined up to addition of pairing of $e_0$ with absolute cohomology class in $X$ (i.e. up to addition of solution to the Picard-Fuchs equation). We will  use this freedom to get rid of omitted terms (we will see later that this is possible).
  
   The potential is in general multivalued, therefore we introduce new coordinates ${\tilde q}^1, ...,{\tilde q}^r$ in such a way that the potential is a one-valued function in these coordinates (see a precise definition below).
We can rewrite the expression for the normalized potential
in the form
 \begin{equation}
\label{W}
\tilde {\mathcal   
  W}=\sum _{{\bf k}}\sum _{d\in \mathbb{N}}{d}^{-2}N_{\bf k}{\tilde q}^{d\bf k} 
\end{equation}
{\it We  will prove that (up to a constant factor) the coefficients $N_{\bf k}$ are integers. }This will be consistent with the interpretation of these numbers in terms of holomorphic disks in mirror A-model. The constant factor in this statement
depends on the choice of the element of  relative cohomology group used in the definition of the potential. 

  More precisely, one should consider a family $\tilde\mathcal{{ M}}$ of  pairs $(X,Y)$ where $X$ is a Calabi-Yau threefold and $Y$ a one-dimensional subvariety ; we assume that this family is defined over $\mathbb{Z}$  and the coordinates on it agree with this structure.  There exists a natural projection $p$ of the family $\tilde \mathcal{{ M}}$ into  $\mathcal{M}$ ;
we assume that this projection is a covering map with finite fibers.  In particular, the family $\tilde \mathcal{{ M}}$ has the same dimension as  $\mathcal{M}.$ The relative cohomology $H^3(X,Y)$ specifies a vector bundle over $\tilde\mathcal{{ M}}$ ; this bundle is equipped with a flat connection $\nabla$ (Gauss-Manin connection) that is compatible with Gauss-Manin connection on absolute cohomology.  {\footnote {Notice, that in compact case
   instead of relative cohomology one
 can consider de Rham cohomology of $X\setminus Y$ with compact support . Using Poincare duality one can identify the relative homology $H_i(X,Y)$ with the de Rham cohomology  $H^{6-i}(X\setminus Y)$. }}
 Relative homology $H_3(X,Y)$  is dual to the relative cohomology; it 
  is also equipped with Gauss-Manin connection.
  We have exact sequences
  \begin{equation}
\label{e}
 0\to H_3(X)\to H_3(X,Y)\to Ker (H_2(Y)\to H_2(X))\to 0,
\end{equation}
 \begin{equation}
\label{ee}
  0\to H^2(Y)/Im H^2 (X)\to H^3(X,Y)\to H^3(X)\to 0. 
\end{equation}  
  We will start with the simplest situation when the algebraic Chern class $c_2$ is represented by a  subvariety $Y$ having two irreducible components; more precisely, we assume that $Y =Y_1\bigcup Y_2$ and $c_2$ is represented by a homologically trivial linear combination of $Y_1$ and $Y_2$. 
(If $X$ is a projective variety a homologically trivial subvariety  $Y$ cannot be irreducible.)
 In this case
  the group $H^2(Y)/Im H^2(X)$ is one-dimensional, hence 
  $\dim H^3(X,Y)=\dim H^3(X)+1.$ Notice that (\ref {e}) and (\ref {ee}) remain valid for any group of coefficients; they are valid also for de Rham cohomology over any group (in particular, over integers) because the families at hand are defined over $\mathbb{Z}$.  We will work with cohomology of $X\setminus Y$ that is Poincare dual to relative homology;
  in these terms the exact sequence (\ref {e}) leads to exact sequence
  \begin{equation}
\label{E}
0\to H^3(X)\to E \to I \to 0
\end{equation}
where $E=H^3(X\setminus Y)$ and $I$ is one-dimensional.  It is important for us that one can consider the sequence (\ref {E}) also for the cohomology with coefficients in $\mathbb{Z}_p$
and that this sequence is compatible with Frobenius map. (Notice, however, that in $p$-adic situation one can identify $E$ with $H^3(X\setminus Y)$  only in the case when $Y$ is smooth.)
{\footnote  {On $H^3(X\setminus Y) $ we have  the Hodge filtration and the weight filtration as in cohomology of any algebraic variety . 
These data specify the mixed Hodge structure in the cohomology. (More precisely, one should talk about variation of Hodge structure.) This mixed structure can be considered as an extension of  one-dimensional pure Hodge structure $\mathbb{Z}(-2)$ by a pure Hodge structure on $H^3(X)$ .This  means, in particular, that
in the Hodge filtration on $I$ the terms $F_k$ with $k\leq 2$ coincide with $I$ and all other terms vanish.  The extension at hand corresponds to normal function used in \cite {MW} to describe the potential.}} 
    
  Let us assume that we are  working in the
  neighborhood of the boundary point of $\mathcal{M}$ with maximally unipotent  
  monodromy. Then we can use the basis $e_0, e_a, e^a, e^0$ in absolute three-dimensional cohomology where the Gauss-Manin connection has the form (\ref {gm}) (in particular, $e^0$ is  a covariantly constant holomorphic section). The images of the vectors of this basis in $H^3(X\setminus Y)$ will be denoted by  ${\tilde e}_0, {\tilde e}_a, {\tilde e}^a, {\tilde e}^0$. These vectors
 together with vector $B$ constructed as a preimage of fundamental class $b$
in $I$ 
form a basis in  $H^3(X\setminus Y)$. {\footnote {We are choosing the class $b$ imposing the condition of algebraic integrality. To compare our results with the results of \cite {MW} one should 
normalize this class imposing the condition of topological integrality. This means that we should multiply $b$ by $(2\pi i)^{-2}$. The relation 
between topological and algebraic integrality can be derived from the fact $I$ comes from two-dinensional cohomology of variety $Y$ having complex dimension 1. (In general, if we have an $n$-dimensional variety over $\mathbb{Z}$ then in its $k$-dimensional cohomology with coefficients in $\mathbb{C}$ we have a lattice of
algebraically integral elements and a lattice of topologically integral elements. If $k=2n$ the first lattice can be obtained from the second one by means of multiplication by $(2\pi i)^{n}$.)}}

Of course, there is some freedom in the choice of $B$, however, we can impose conditions that  eliminate this freedom. Namely, we will prove in  Appendix that there is a unique lifting $B$ of $b$ obeying $B\in F^2$ and 
\begin{equation}
\label{b}
\nabla _k B=\tau _{ak}{\tilde e}^a.
\end{equation} 
The vector $B$ satisfying these conditions is algebraically integral; together with ${\tilde e}_0, {\tilde e}_a, {\tilde e}^a, {\tilde e}^0$ it 
generates  de Rham cohomology
$H^3(X\setminus Y)$ with coefficients in $\mathbb{Z}.$  {\footnote {More precisely, we  prove these statements for the cohomology over $\mathbb{Z}_p.$ If we were able prove them for all primes the statements over $\mathbb{Z}$ would follow.  However, we exclude $p=2,3$ and require that the reduction of our family mod $p$ is smooth.}}

The number of holomorphic disks can be 
expressed in terms of $ \tau _{ak}$. To check this we notice that covariantly constant
elements of  $H^3(X\setminus Y)$  can be expressed in these terms. (We are talking about covariantly constant elements that are Poincare dual to non-trivial relative cycles, i.e. to cycles that do not come from absolute cycles in $X$.) Namely, we can find a covariantly constant element $T$  (flat section) in the form 
\begin{equation}
\label{bb}
T=B-\alpha_{0}{\tilde e}^0-\alpha _{a}{\tilde e}^a
\end{equation}
where 
\begin{equation}
\label{a}
\delta _k\alpha_{0}+\alpha _k=0,\\
\delta _k \alpha _a=\tau _{ak}.
\end{equation}{\footnote {The coefficients $\alpha_{0},\alpha_{k}$ specify a normal function,
the first of  equations (\ref {a}) coincides with transversality condition for normal function.}}
We are working in canonical coordinates on $ \mathcal{{ M}}$ and using the fact that in these coordinates $\nabla _ke^a=\delta _k^ae^0$. 

Strictly speaking we should work on $\tilde \mathcal{{ M}}$ where canonical coordinates are 
not genuine coordinates; holomorphic functions on this space are multivalued functions of canonical coordinates on $ \mathcal{{ M}}$ . It  is better to introduce
an analog of canonical coordinates on $\tilde \mathcal{{ M}}$. They can be defined as  coordinates ${\tilde q}^1, ..., {\tilde q}^r$ compatible with integral structure 
on $\tilde \mathcal{{ M}}$ and related to canonical coordinates on $ \mathcal{{ M}}$
by the formula $q^a= 
{\tilde q}^{{\bf s}^a}$, or,more precisely,
$$q^a=({\tilde q}^1)^{s^a_1}...({\tilde q}^r)^{s^a_r}.$$
Here $s^a_j$ are non-negative integers.  We assume that such coordinates exist.

For quintic $q={\tilde q}^2$.

Notice, that there is some freedom  in the definition of $T$ (we can add ${\tilde e}^0$ multiplied by any constant). We will use this freedom to impose condition that $\alpha _0$ vanishes at the boundary point.  From the other side calculating the pairing between $T$ and ${\tilde e}_0$ we see that  $\alpha _0$ can be interpreted as
 an integral of $e_0$ over a non-trivial relative cycle. This means that it  can be considered  as the normalized potential $\tilde \mathcal{W}$, therefore the number of holomorphic disks can be expressed in terms of  $\alpha _0$. (Notice, $T$
 has no logarithmic terms in the neighborhood of the boundary point, therefore we have justified  the statement about possibility to eliminate such terms in the normalized potential.)
 
The basis we
consider is  algebraically rational (moreover, it is integer ), hence  we can
consider $p$-adic reduction of our construction.
The  homomorphisms of the exact sequence (\ref {E}) commute with
Frobenius map. Applying this fact and taking into account that ${\rm Fr}b=p^2 b$ we obtain 
$${\rm Fr}B=p^2 B+\mu _0 {\tilde e}^0+\mu _a {\tilde e}^a.$$ (Writing this formula we have used also that Frobenius map is compatible with monodromy weight filtration; see Appendix.)
Using (\ref {gmf}) we see that
\begin{equation}
\label{n}
\delta _k\mu _0+\mu _k=0,
\end{equation}
\begin{equation}
\label{nn}
p^2\tau _{ak}+\delta _k \mu _a=p^2{\rm Fr}\tau _{ak}.
\end{equation}
Now we can use (\ref {a}) to obtain from  (\ref {nn}) that
$$p^2\delta _k\alpha _a+\delta _k \mu _a=p\delta _k{\rm Fr}\alpha _a,$$
hence 
$$p^2\alpha _a+\mu _a=p{\rm Fr} \alpha _a +c_a,$$
where $c_a$ are constants. Applying this formula and (\ref {a}) to (\ref {n}) we see that
$$p^2\delta _k\alpha _0+\delta _k \mu _0+c_k=\delta _k{\rm Fr}\alpha _0.$$
Taking into account that all terms containing logarithmic derivatives vanish at the boundary point $q=0$ we see that $c_k=0$. Integrating the last equation we obtain
\begin{equation}
\label{f}
{\rm Fr} \alpha _0-p^2 \alpha _0=\mu _0 +\gamma.
\end{equation}
Recall that we have assumed that  $ \alpha _0$
vanishes at the point ${\bf q}=0$, hence  $\mu _0 +\gamma$ also vanishes at this point.  We will work in canonical coordinates ${\tilde q}^1, ...,{\tilde q}^r$ on $\tilde \mathcal{{ M}}$. Then  all coefficients of the power expansion of $\mu _0+\gamma$ are divisible by $p^2$ (the constant term vanishes; the divisibility of other coefficients follows from divisibility properties of Frobenius map).

Now we can apply  Lemma 3$'$ to express $\alpha _0$ in terms of $\mu _0$. Using this expression and the fact that $\mu _0+\gamma     $ is divisible by $p^2$
we see that 
$$\alpha _0=\sum_{\bf k}\sum_{d\in \mathbb{N}}\frac{1}{d^2}N_{\bf k}{\tilde q}^{d{\bf k}}=
\sum_{\bf k}N_{\bf k}{\rm Li}_2({\tilde q}^{\bf k}),$$
where $N_{\bf k}$ are integers. These integers represent the numbers of holomorphic disks.

Notice that we have constructed the flat section
$T$ starting with algebraically integral generator $b$ of $I$.
To reproduce the results of \cite {MW}  we should take as $b$ a topologically integral generator of $I$. This leads to the appearance of
a constant factor $(2\pi i)^{-2}$ in the expression
of  the potential in terms of holomorphic disks.
(See the footnote $^{[18]}$ .)

The consideration in the case when the variety $Y$ has more components is very similar. In this case we still can construct  a one-dimensional extension $E$ of $H^3(X)$ that is included in
exact sequence 
\begin{equation}
\label{nc}
0\to H^3(X)\to E \to I=\mathbb{Z}(-2)\to 0,
\end{equation}
that generalizes the sequence (\ref {E}). Again $E$ is equipped with mixed Hodge structure that  can be considered as an extension of two pure Hodge structures. (More precisely, we should talk about family of Hodge structures equipped with Gauss-Manin connection or about variation of Hodge structure.) More details about this picture can be found in \cite {AM}, \cite {KLM},
\cite {Gr}.

The above calculations can be repeated without modifications in this more general situation.

The  cases when we have open string moduli or/and  Calabi-Yau manifold $X$ is not compact are similar, but more complicated.
Let us consider, for example, the situation analyzed in  \cite {L}, \cite {L1}, \cite {L2}, \cite {L3}. It was shown in these papers that under certain conditions the number of holomorphic disks can be expressed in terms of connection
on appropriate cohomology bundle  over the moduli space $\mathcal{M}$ of pairs $(X, B)$ where $X$ is a Calabi-Yau manifold and $B$ belongs to a family of $B$-branes . Namely, the authors construct what they call $ N=1$ special geometry of this bundle. {\footnote { The authors of these papers consider the case of family of $B$-branes corresponding to a family of one-dimensional subvarieties $B$ with trivial vector bundle. Then the cohomology bundle is the bundle of middle-dimensional relative cohomology groups). However, our discussion is more general;
it relies only on the properties of connection associated with $ N=1$ special geometry. }} From the mathematical viewpoint one can say that that they construct a holomorphic basis $e_0, e_a,e_{\alpha}, ... $
in the fibers of the cohomology bundle in such a way that Gauss-Manin connection has the
form
$$\nabla _ae_0=Y^b_ae_b,$$
$$\nabla _ae_b=c^{\alpha}_{ab}e_{\alpha},$$
$$\nabla _a e_{\alpha}=...$$
Here  $e_0$ corresponds to  the normalized Calabi-Yau form, $...$ stands for the elements of basis 
that do not play any role in our calculations and 
for their linear combinations.
 For the dual basis
$e^0, e^a,e^{\alpha},...$ the Gauss-Manin connection obeys
$$\nabla _ae^0=0,$$
$$\nabla _ae^b=Y^b _a e^0.$$
{\footnote {Notice that  in the situation when $B$-branes correspond to one-dimensional subvarieties the dual cohomology bundle can be interpreted  as
the bundle  of relative homology or, in  compact case, as the bundle of cohomology of the complement $X\setminus B$.}}
We are working in coordinates $z^a$ in the neighborhood of maximally unipotent boundary point $z=0$ of the moduli space $\mathcal{M}$;
the covariant derivative $\nabla _a$ corresponds to the logarithmic derivative with respect to $z^a$.  We assume that the moduli space is defined over integers , the coordinates $z^a$ agree with the integral structure of the moduli space and the Calabi-Yau form is algebraically integral. Then the normalized Calabi-Yau form $e_0$ and its covariant derivatives (in particular, $e_a$)
are also algebraically integral. {\it {We will make stronger assumption that   $e_0, e_a,e_{\alpha}, ... $ generates algebraic de Rham cohomology over $\mathbb{Z}.$}} This assumption can be easily verified in concrete cases.

The condition on the boundary point means that
there exists a basis $g^0,g^a,g^{\alpha}, ...$ consisting of holomorphic (possibly multivalued)
sections of the dual cohomology bundle such
 that $g^0$ is regular at $z=0$, the behavior of $e^a$ is logarithmic, $e^{\alpha}$ behaves as $\log ^2$, etc.  We assume that 
 $$ e^0=g^0, e^a=g^a+\rho ^a g^0,e^{\alpha}=g^{\alpha}+\tau _{\alpha} e^0+\sigma _{\alpha,a} e^a, ...$$ (In other words the dual basis is compatible with the monodromy weight filtration).   
Notice that we do not separate closed string moduli and open string moduli. 

As in the closed string case we introduce the notion of canonical coordinates requiring that  in these coordinates
$Y^a_b=\delta ^a_b$. It is shown in the papers we are using that the number of holomorphic disks can be expressed in terms of the coefficients
$c^{\alpha}_{ab}$ in the formula for Gauss-Manin
connection. These coefficients can be interpreted
as $\delta _a \delta _b \mathcal{W}^{\alpha}$ where $\mathcal{W}^{\alpha} $ stands for the potential  {\footnote {In the situation of   \cite {L}, \cite {L1}, \cite {L2}, \cite {L3} the potential can be defined in terms of integrals of normalized Calabi-Yau form $e_0$  over relative cycles.}} 
that is related with the number of holomorphic disks $n^{\alpha}_\mathbf{ k}$  in the following way:
\begin{equation}
\label{nk}
\mathcal{W}^{\alpha}=\sum n^{\alpha}_\mathbf{ k}{\rm Li}_2(\mathbf{q}^{\mathbf{ k}})
\end{equation}
(Notice that in general the potential can contain
terms with logarithmic behavior at the boundary point. We omit these terms as well as the constant term.)
Here $\mathbf{ q}$ stands for the vector of canonical coordinates, $\mathbf{ k}$ is a multiindex. (Notice that we do not separate closed and open string moduli, therefore the above expression contains not only the numbers of holomorphic disks, but also some combinations of instanton numbers.)
 
 Let us consider  $p$-adic reductions of the cohomology bundle, of the bases in fibers of this bundle and of Gauss-Manin connection. (This is possible because all of these objects are defined over rational numbers.) We will prove that the number of holomorphic disks can be expressed in terms of the matrix of Frobenius operator in canonical coordinates and use this expression to prove integrality of this number. 
 
 First of all we notice that integrality of mirror map in the new situation can be analyzed  in the same way as in Sec. 4. (To prove $p$-integrality we should assume 
 that the matrix $Y^b_a(0)$ is invertible mod $p$.) It follows from the integrality of the mirror map  and from our assumptions 
about $e_0, e_a,
 e_{\alpha},...$ that also in canonical coordinates one can regard $e_0, e_a,
 e_{\alpha},...$ as a generators of de Rham cohomology over $\mathbb{Z}_p$ . Therefore the
 entries of Frobenius matrix in this basis are p-integral, i.e. the coefficients in the formulas
 $$ {\rm Fr}e_0=m^0_0e_0+m^a_0e_a+m^{\alpha}_0e_{\alpha}+...,$$
 $${\rm Fr}e_a=m^b_ae_b+m^{\alpha}_ae_{\alpha}+....$$
 are power series with coefficients from $\mathbb{Z}_p$.
 Moreover, for $p>3$ the coefficients in the first line are divisible by $p^3$, the coefficients in the second line are divisible by $p^2$, etc .(This follows from the divisibility statements for Hodge filtration and from the fact that $e_0$ is the normalized Calabi-Yau form and therefore belongs to $F^3$.  It follows from Griffiths transversality that $e_a=\nabla _a e_0$ belongs to $F^2$, etc.)
 
 We have written the Frobenius matrix in triangular form. To justify this assumption we
 notice that the equation (\ref {sc})  remains valid for the pairing between cohomology bundle and
 the dual cohomology bundle. This means that the Frobenius matrix in the dual basis is obtained by means of transposition and multiplication of matrix entries by some powers of $p$.  The dual basis is compatible with the monodromy weight filtration, therefore the Frobenius matrix in the dual basis (and, hence, in original basis) is triangular. 
   
 As in Sec. 2 we can relate the Frobenius matrix to the Gauss-Manin connection using ( \ref {gmf}). In canonical coordinates we obtain 
 $$\delta_a m^0_0=\delta _a m^b_c=\delta _am^{\alpha}_{\beta}=0,$$
 $$\delta _a m^b_0 +m^0_0\delta ^b_a=pm^b_a,$$
 $$\delta _a m^{\alpha}_0+ m^b_0c^{\alpha}_{ab}=pm^{\alpha}_a,$$
 $$\delta _b m^{\alpha}_a+m^k_ac^{\alpha}_{bk}=pm^{\alpha}_{\beta}{\rm Fr}(c^{\beta}_{ab}.$$
 Here $\delta _a$ stands for logarithmic derivative with respect to the canonical canonical coordinate $q^a.$
 
 We see from the first line that the diagonal entries of the Frobenius matrix are constant
; using other equations at $z=0$ we can express all of them in terms of $m^0_0$ . We obtain $m^a_b=p^{-1}m^0_0\delta ^a_b, m^{\alpha}_{\beta}=
p^{-2}m^0_0 \delta ^{\alpha} _{\beta}$ (we used the fact that at $z=0$ the logarithmic derivative vanishes). It follows from the second line that $\delta _am^b_0$ is a constant; taking $z=0$ we see that this constant is equal to zero. 
We see that $m^b_0$ itself is a constant. One can prove that this constant vanishes. We obtain that  $$\delta _a m^{\alpha}_0
=pm^{\alpha}_a,$$
hence
$$\delta _a\delta _b m^{\alpha}+ m^0_0c^{\alpha}_{ab}=m^{0}_{0}{\rm Fr}(c^{\alpha}_{ab}).$$
One can prove that $m^0_0=\pm p^3$. (This follows from (\ref {cs}) and from the fact that $e^0$ is an eigenvector of  Fr with eigenvalue $\pm 1$.)  Using the equation
$c^{\alpha}_{ab}=\delta _a \delta _b \mathcal{W}^{\alpha}$ we obtain
\begin{equation}
\label{w}
p^{-2}{\rm Fr}\mathcal{W}^{\alpha}-\mathcal{W}^{\alpha}=\pm p^{-3}m^{\alpha}+l^{\alpha}
\end{equation}
where $l^{\alpha}$ is an (inhomogeneous) linear function with respect to $\log q^a$.
Using Lemma 3$'$ we obtain an expression of 
numbers $n^{\alpha}_{\mathbf{ k}}$ defined in (\ref {nk}) in terms of RHS of equation (\ref {w}).
If $p>3$ we can say that $m^{\alpha}$ is divisible by $p^3$. In the case when $l^{\alpha}=0$ this guarantees the integrality of the RHS and,therefore, the integrality of  $n^{\alpha}_{\mathbf{ k}}$.  Notice that in the case when the potential is purely nonperturbative (i.e. it does not contain constant and logarithmic terms) the
condition  $l^{\alpha}=0$ is satisfied. 

{\bf Appendix}

We would like to look at the above constructions from more mathematical viewpoint.

	Recall, that we take as a starting point a  smooth miniversal family of Calabi-Yau threefolds  with the base $\mathcal{M}$. We assume that this family is defined over complex numbers and can be extended to (non-smooth)  semistable  family {\footnote {One says that the family is semistable if the singular fibers have only mild singularities. For example,if $\mathcal{M}$ is one-dimensional a singular fiber should be a normal crossing divisor in the total space. In general the preimage of $\tilde {\mathcal{M}}\setminus \mathcal{M}$ in the total space should be a normal crossing divisor.}}with a base $\tilde{\mathcal{M}}$. We are working in a neighborhood of a  regular  boundary point $z=0$. In the neighborhood of this point we can introduce coordinates $z^1,...,z^r$ in such a way that  the complement of
$\mathcal{M}$ in $\tilde{\mathcal{M}}$ is a normal crossing divisor $z^1...z^r=0$.  In our situation one can prove  that the bundle of middle-dimensional cohomology $\mathcal{H}$ and Hodge filtration on this bundle can be extended to $\tilde{\mathcal{M}}$  so that the Gauss-Manin connection has  first order pole with nilpotent residues $N_i$. (We consider here
cohomology with coefficients in $\mathbb{C} $. )These residues can be considered as operators action on the special fiber $\mathcal{F}$  (i.e. on the fiber over $z=0$) of the cohomology bundle  $\mathcal{H}$.  We construct an operator $N$ as a linear combination of $N_i$ with positive coefficients. As every nilpotent operator the operator $N$ induces an increasing filtration $V_0\subset V_1\subset ...$ specified by the conditions $a)$ $N(V_k)\subset V_{k-2}$, $b)$ $N^k$ generates an isomorphism of $V_{n+k}/V_{n+k-1}$ with $V_{n-k}/V_{n-k-1}$ (weight filtration).  Here $n$ denotes the order of nilpotent operator ( the maximal number obeying $N^n\neq 0$.) In our case $n=3$, $V_0= Im N^3,...$ One can prove that the weight filtration does not depend on the choice of linear combination
used in the construction of $N$.

We shall impose a condition that  there exists  a   vector  $e_0$ such that $\mathcal{F}$ is generated by vectors obtained from $e_0$ by successive action  of operators $N_i$.  Then one can prove that $V_{2k}=V_{2k+1}$ ; to  reconcile the
above notations with the notations in the main body of the paper we shall write $W_k=V_{2k}.$ It is easy to check that $\dim W_0=\dim V_0=\dim V_1=1,\dim W_1= \dim V_2=\dim V_3=1+r$, $ \dim W_2=\dim V_4=\dim V_5=1+2r, \dim W_3=\dim V_6=2+2r.$  This means  that the boundary point is maximally unipotent .  We  always can choose the vector $e_0$ to belong to $F^3$ (to the smallest element of Hodge filtration $F^k$). This condition specifies the vector uniquely (up to a constant factor); we impose it in what follows. Let us denote by $\mathcal{ F}_s$ the subspace spanned by vectors obtained by means  of application of $s$ operators $N_i$ to $e_0$. One can check that $W_k$ is a direct sum of $\mathcal{F}_s$ with $s\geq 3-k$  and $F^k$ is a direct sum of $\mathcal{F}_s$ with $s \leq 3-k.$ (Recall that  $N_i$ comes from Gauss-Manin connection, hence it follows from Griffiths transversality that $N_i$ acts  from $F^k$ into $F^{k-1}$.) It follows from the above statements that $\mathcal{F}$ is a direct sum of  intersections 
$W_k\bigcap F^k=\mathcal{F}_{3-k}$ (this means that the mixed Hodge structure on the special fiber is  a Hodge-Tate structure). 

In our definitions we had assumed that our family is defined over $\mathbb{C}$ and  had in mind  cohomology groups with coefficients in complex numbers. Now we can consider families over other rings and  de Rham cohomology with other groups of coefficients. We will assume that the cohomology with coefficients in $\mathbb{Z}$ has no torsion.  For simplicity of exposition we assume that our family is defined over $\mathbb{Z}$ and it is smooth over $\mathbb{Z}$; this means that it it remains smooth after reduction mod $p$ for all prime numbers. The last assumption is not very realistic; usually the assumption of  smooth reduction is violated  for finite number of "bad" primes. Then the family is smooth over $\mathbb{Z}[N^{-1}] $ where $N$ is a product of bad primes (i.e. over the ring of fractions with denominators containing only bad primes) and our statements can be generalized to this case. We will  consider the cohomology with coefficients in a torsion-free group; then the above picture of the cohomology of the special fiber does not change if we  require that  the cohomology is generated
by elements obtained from one vector $e_0$ by means of successive application of  operators $N_i$. {\footnote {Notice that for cohomology with coefficients in $\mathbb{Z}_p$
this condition is equivalent to the condition on the behavior of Yukawa couplings mod $p$ at the boundary point imposed in Sec. 2 }}

{\it We assume that this condition is satisfied by the cohomology with coefficients in $\mathbb{Z}_p$  if $p>3$  and the mod $p$ reduction is smooth.} 
Our main tool is the Frobenius map ${\rm Fr}$ on the  $\mathbb{Z}_p$-cohomology that exists if the mod $p$ reduction of our family is semistable. 
Let us denote by $\alpha$ the map of the base transforming a point with coordinates $z^1,...,z^r$ into a point with coordinates $(z^1)^p,...,(z^r)^p$. Then ${\rm Fr}$ can be considered as a map
from  a bundle $\alpha ^*\mathcal{H}$ into a bundle $\mathcal{H}$ where $\mathcal{H}$ is understood as a bundle of  $\mathbb{Z}_p$-cohomology; it induces a map of sections, also denoted by ${\rm Fr}$.
{\footnote {This statement is equivalent to the relation (\ref {fr}) in Sec. 2.}} 

 From the relations of ${\rm Fr}$  with Gauss-Manin connection and scalar product (see  (4) and (5) in Sec. 2)  we obtain that  on the special fiber
$$ N_i{\rm Fr}=p{\rm Fr}N_i$$,
$$<{\rm Fr}x,{\rm Fr}y>=p^3<x,y>.$$
Making use of the first of these relations we can
express the action of ${\rm Fr}$ on the special fiber in terms of the  action of ${\rm Fr}$ on $e_0$. We obtain that  ${\rm Fr}$ preserves the filtration $W_k$ and that on $W_k/W_{k-1}$ it acts as multiplication by a number  that we will denote   $\lambda _k$.
These numbers are related by the formula
$\lambda _k=p\lambda _{k-1}$, hence $\lambda _3=p^3 \lambda _0.$
From the second relation we obtain
$\lambda _0 \lambda _3=p^3$,
hence 
$$\lambda _0 =\pm 1.$$
We will assume that $\lambda _0=1$; the modifications necessary in the case $\lambda _0=-1$ are obvious.

One can prove that  ${\rm Fr}x$ for $x\in F^k$  is divisible by $p^k$. (If $k=1$ this is true for all $p$; if $k>2$ one needs an additional assumption 
$p>3$.)

One can prove the following theorem that is instrumental in the derivation of our main results. 

{\bf Theorem} 

The filtration $W_0\subset W_1\subset W_2\subset W_3$ on the special fiber can be extended to the filtration on the cohomology bundle (monodromy weight filtration) that is compatible with Gauss-Manin connection. The 
monodromy weight filtration is opposite to the Hodge filtration $F^3\subset F^2\subset F^1\subset F^0.$ (In other words the cohomology is a direct sum of intersections $ W_k\bigcap F^k$.)

This theorem is well known for cohomology with coefficients in $\mathbb{C}$. {\footnote {To construct the monodromy weight filtration in this case one uses the weight filtration corresponding to the logarithms $ \tilde{N}_j$ of monodromy transformations (or, more precisely to their linear combination). This filtration obeys the conditions of the theorem because at the boundary point $ \tilde{N}_j=2\pi i N_j$.}} We will sketch a proof of it   in much more difficult case of cohomology with coefficients in $\mathbb{Z}_p$, assuming that the reduction of our family mod $p$ is again a smooth family and that $p>3$.
{\footnote {Notice that it follows from this statement that the theorem is correct for any ring where 2,3 and all prime numbers with non-smooth reduction are invertible.}}

Let us start with the construction of the extension of  $W_0$. We will take any section $x$ of $\mathcal{H}$ defining an element in the special fiber that is invariant with respect to ${\rm Fr}$ . It is easy to prove that the sequence ${\rm Fr}^nx$ converges and the limit is a flat (covariantly constant) section. (This can be derived from the fact that  the covariant derivative of ${\rm Fr}^nx$ is divisible by $p^n$; this fact follows from (4).) We obtain that for every element of special fiber belonging to $W_0$ there exists a unique flat section of $\mathcal{H}$ containing this element.
The set of such sections specifies an extension of $W_0$; we will use the same symbol for this extension. Notice that $\mathcal{H}$ is a direct sum of $W_0$ and $F^1$; this follows from the corresponding fact in special fiber. {\footnote {Flat sections constructed above correspond to holomorphic solutions to the Picard-Fuchs equation. We obtain from this remark the integrality property  of these solutions.}}

As we have seen the action of Frobenius map on the special fiber induces a multiplication by $p$ on $W_1/W_0$.  We factorize $\mathcal{H}$ with respect to the extension of $W_0$; the quotient is isomorphic to $F^1$, hence a Frobenius map 
on $\mathcal{H}/W_0$ is divisible by $p$. We can apply  the above  consideration to ${\rm Fr}/p$
to construct for an arbitrary element
of $W_1/W_0$ a unique flat section of
$\mathcal{H}/W_0$  that contains  this element.
These sections specify an extension  of $W_1/W_0$; lifting them (non-uniquely) to $\mathcal{H}$ we obtain an extension of $W_1$ (again denoted by the same symbol). It is obvious that
this extension agrees with Gauss-Manin connection. 

Applying this construction again we obtain the proof of the theorem.

Notice that it is obvious from the construction that the monodromy weight filtration is invariant with respect to Frobenius transformation.

The above theorem can be applied to construct the basis  $e^0,e^a,e_a,e_0$ that we used in the main body of the paper. In particular, we take $e^0$ as a generator of $W_0$. To construct $e^a$ we take a free system of generators of
$W_1/W_0$ and lift it to $W_1\bigcap F^1$.   To construct other vectors of symplectic basis $e^0,e^a,e_a,e_0$ we use the fact that the Poincare pairing determines  a non-degenerate pairing  between  $W_k\bigcap F^k=\mathcal{F}_{3-k}$  and $W_{3-k}\bigcap F^{3-k}=\mathcal{F}_{k}.$ (This fact is correct for any  ring of coefficients.)

The matrix of Frobenius map is triangular in the
basis $e^0,e^a,e_a,e_0$; this follows from the
compatibity of Frobenius map with monodromy weight filtration. In particular,
$${\rm Fr}e^0=\pm e^0,$$
$${\rm Fr}e^a=\pm p( e^a+m^ae^0),$$
where $m^a$ is a constant (as follows from (\ref {gmf})). Using the theory of motives one can prove that that this constant is an integer. We assume that this constant is zero (we always can change the vector $e^a$ adding an integer multiple of $e^0$ to satisfy this condition.)

Let us discuss now the construction of the
basis ${\tilde e}_0, {\tilde e}_a, {\tilde e}^a, {\tilde e}^0, B$ of  $E$ that we used in the last section. 
The first vectors of this basis come from  $H^3(X)$, hence we should construct only the vector $B$. We start with the construction of this vector in the special fiber; {\footnote {The exact sequence (\ref {E})  or (\ref {nc}) can be extended to 
the cohomology bundle over $\tilde \mathcal{M}$ and,
in particular, to the special fiber.}}  to do this we are using the exact sequence (\ref {E}) to lift the generator $b$ of $I$
to a vector $B\in F^2\subset  H^3(X\setminus Y)$
such that  $N_kB  $  is a linear combinations of vectors ${\tilde e}^a.$ To give a more invariant definition of $B$ we can use the
weight filtration $V_k$ corresponding to linear combination $N$ of operators $N_k$ constructed as residues of Gauss-Manin connection.  One can check that  rank $V_0$= rank  $V_1=1$,
rank $V_2/V_1=r$, rank $V_3/V_2=1$, rank $V_4/V_3=r$, rank $V_5/V_4=1$, rank $V_6/V_5=0.$
The special fiber is a direct sum of $V_0\bigcap F^0$, $V_2\bigcap F^1$, $ V_4 \bigcap F^2$ and $V_5\bigcap F^3.$ (This means that skipping  some terms of weight filtration we obtain a Hodge-Tate structure on the special fiber.) It follows that the natural map of $V_3\bigcap F^2$
onto $V_3/V_2$ is an isomorphism. The generator $b$ of $I$ corresponds to a generator
of $V_3/V_2$ and this generator can be lifted in unique way to the generator $B$ of $V_3\bigcap F^2$. It follows from Griffiths transversality and 
from the fact that $b$ is covariantly constant that
$N_k B \in V_2\bigcap F^1.$ This means that
$N_kB  $  is a linear combinations of vectors ${\tilde e}^a.$

Let us consider the cohomology with coefficients in $\mathbb{Z}_p$ assuming that the reduction mod $p$ gives a semistable family. (Until now we did not fix the coefficients of cohomology.) The exact sequence (\ref {E}) is compatible with Gauss-Manin connection, Hodge filtration and  Frobenius map. Using these facts we can check that the filtration $V_k$ on the special fiber can be extended to  
the filtration on the cohomology bundle (monodromy weight filtration) that is compatible with Gauss-Manin connection (we use the same symbol $V_k$ for the extension).  To prove this we repeat the arguments used in the proof of the above theorem (we construct the extensions inductively, starting with the extension of  the
quotient $V_k/V_{k-1}$ ). Extending $V_3/V_2$
we use the fact that Fr acts on this space as a multiplication by $p^2$.  Notice that lifting the
covariantly constant section of the extended $V_3/V_2$ we obtain a unique section $B\in V_3\bigcap F^2$ extending the vector $B$ in the special fiber. By construction the monodromy weight filtration is compatible with Frobenius map. The sets $V_0\subset V_2\subset V_4\subset V_6$ together with Hodge filtration 
specify Hodge-Tate structure on the cohomology bundle. This means that the cohomology is a direct sum of $V_0\bigcap F^0$ (spanned by ${\tilde e}^0)$, $V_2\bigcap F^1$ (spanned by ${\tilde e}^a)$, $V_4\bigcap F^2$ (spanned by ${\tilde e}_a$ and $B$ ) and $V_6\bigcap F^3$(spanned by ${\tilde e}_0.$)

{\bf Acknowledgments.} We are indebted to  M. Aganagic, M.Aldi,
D. Kazhdan, M. Kontsevich, A. Ogus, I. Shapiro, J.Walcher  for useful discussions.

 \end{document}